\documentclass[%
 preprint, 
nofootinbib,
 amsmath,amssymb,
 aps, physrev,
]{revtex4-2}

\usepackage{epigraph}
\usepackage{graphicx}
\usepackage{dcolumn}
\usepackage{bm}
\usepackage{caption}

\usepackage{todonotes}

\usepackage[normalem]{ulem}
\usepackage{xcolor}

\begin{document}


\title{\textbf{The Pareto Frontiers of Magic and Entanglement:\\ The Case of Two Qubits} 
}%

\author{Alexander Roman}
\author{Marco Knipfer}
\author{Jogi Suda Neto}
 \altaffiliation[Also at ]{Quantum Theory Institute, CERN, Geneva, Switzerland.}
\author{Konstantin T. Matchev}
 \email{Corresponding author: kmatchev@ua.edu}
\author{Katia Matcheva}
\author{Sergei~Gleyzer}
\affiliation{%
Department of Physics and Astronomy, University of Alabama, Tuscaloosa, AL USA
}%



\date{\today}

\begin{abstract}
Magic and entanglement are two measures that are widely used to characterize quantum resources. We study the interplay between magic and entanglement in two qubit systems, focusing on the two extremes: maximal magic and minimal magic for a given level of entanglement. We quantify magic by the R\'{e}nyi entropy of order 2, $M_2$, and entanglement by the concurrence $\Delta$. We find that the Pareto frontier of {\em maximal} magic $M_2^{\text{(max)}}(\Delta)$ is composed of three separate segments, while the boundary of {\em minimal} magic $M_2^{\text{(min)}}(\Delta)$ is a single continuous line. We derive simple analytical formulas for all these four cases, and explicitly parametrize all distinct quantum states of maximal or minimal magic at a given level of entanglement.
\end{abstract}

\maketitle
\newpage

\tableofcontents


\newpage
\section{Introduction}
\label{sec:intro}

Entanglement and magic are important and useful resources in quantum computing \cite{Chitambar:2018rnj}.
Entanglement is a uniquely quantum effect that lacks any classical equivalent; it serves as a cornerstone for the performance of most quantum algorithms. However, according to the Gottesman-Knill theorem \cite{Gottesman:1998hu,Gottesman:1999tea,Aaronson:2004xuh}, entanglement by itself does not ensure a quantum advantage, as certain classes of highly entangled states can still be efficiently simulated by classical means. This is why, to quantify the computational overhead of simulating specific quantum states on classical hardware, Braviy and Kitaev introduced the concept of ``magic'' (also known as non-stabilizerness) \cite{Bravyi:2004isx}. It is believed that simulating {\em maximally magic} states is exponentially difficult using classical resources, yet magic states are required for universal quantum computing. Therefore, there has been a lot of recent interest in identifying and characterizing the states with maximal magic in various quantum systems from the point of view of quantum resource theory \cite{Liu:2020yso,Wang:2023uog,Cuffaro:2024wet,Liu:2025frx,Ohta:2025utz}. In fact, the study of magic is no longer confined to quantum algorithms; it is now gaining traction as a vital tool for understanding phenomena in many diverse areas like 
quantum field theory \cite{White:2020zoz,Nystrom:2024oeq,Cepollaro:2024sod},
particle physics \cite{White:2024nuc,White:2024bjp,Liu:2025qfl,Aoude:2025jzc,Busoni:2025dns,Gargalionis:2025iqs,Liu:2025bgw,Durupt:2025wuk,Cheng:2025zaw,Robin:2025ymq,Nunez:2025xds,Hatta:2025obw,Altakach:2026fpl,Esposito:2024uzw,Guo:2026yhz,Afik:2026pxv,Gargalionis:2026onv},
condensed matter physics \cite{Ellison:2020dkj,Oliviero:2022euv,Sierant:2025fax},
nuclear physics \cite{Robin:2024oqc,Brokemeier:2024lhq}, 
astrophysics \cite{Leone:2022afi,Yang:2025zrl},
and many others.

\begin{figure}[t]
    \centering
    \includegraphics[width=0.8\linewidth]{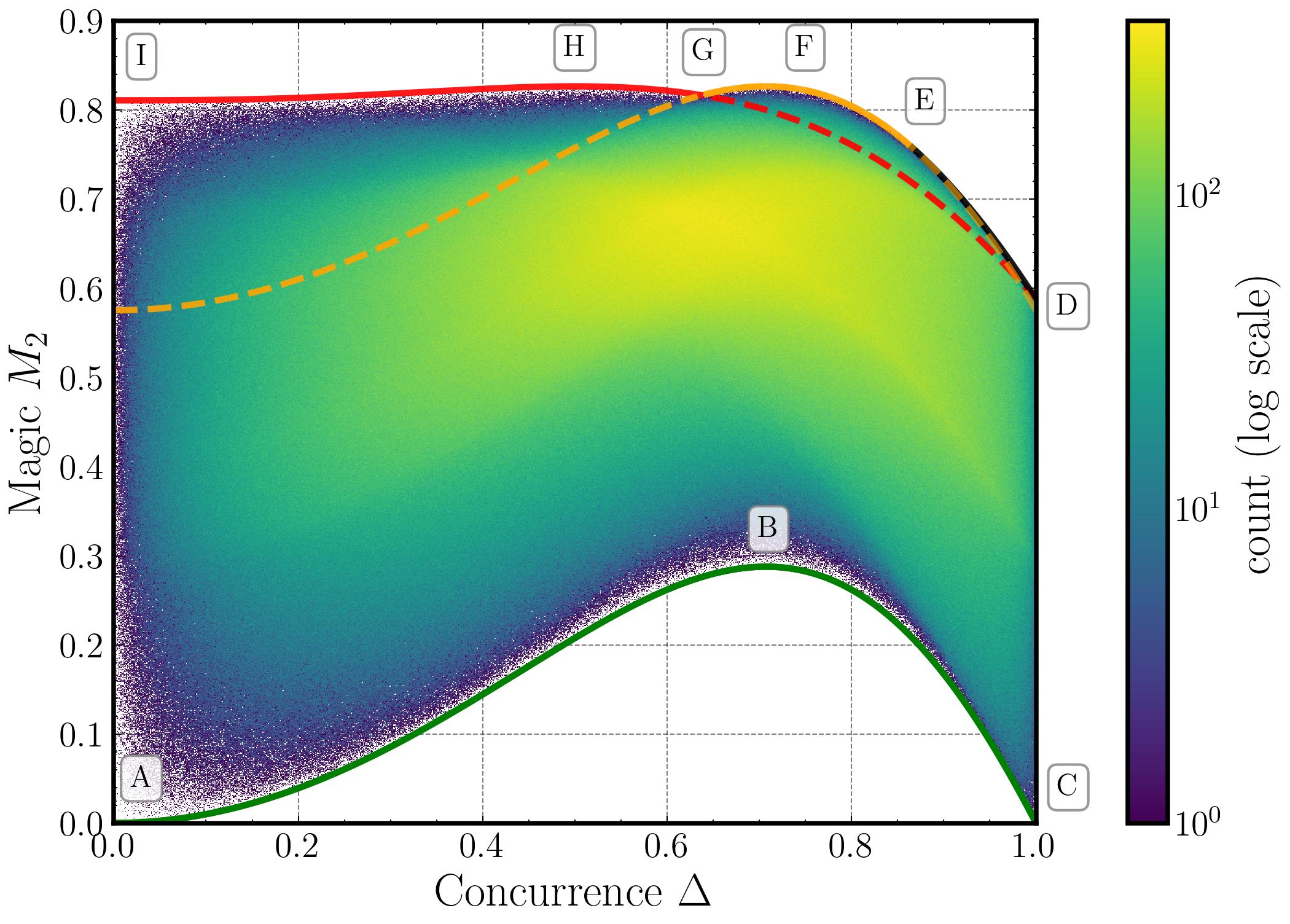}
    \caption{
    2D histogram of the $|\psi\rangle$ number density in the $(\Delta, M_2)$ plane. The distribution was generated by sampling 50 million $|\psi\rangle$ states according to the Haar measure\footnote{The Haar measure is the unique unitarily invariant measure, providing a uniform distribution over pure states~\cite{Mezzadri:2006nac}.} and binned into a $1000 \times 1000$ grid. 
    Points A through I mark interesting features discussed in the text. Colored lines represent the analytic predictions from eqs.~(\ref{eq:M2min_delta}-\ref{eq:f_functions}) derived in this paper. Dashed lines are extensions of those lines into the bulk. }
    \label{fig:M2_versus_Delta}
\end{figure}

In this paper, we study the interplay of entanglement and magic in the general state $\vert \psi\rangle$ of two qubits. We choose to quantify the entanglement of $\vert \psi\rangle$ in terms of the concurrence~$\Delta$~\cite{Hill:1997pfa,Wootters:1997id} defined in Section~\ref{sec:entanglement}, while the magic of $\vert \psi\rangle$ will be measured in terms of the R\'enyi-2 entropy~$M_2$~\cite{Leone:2021rzd} defined in Section~\ref{sec:magic}. To set up the stage for the analysis in this paper, in Fig.~\ref{fig:M2_versus_Delta} we show a high-resolution 2D histogram of the $\vert \psi\rangle$ number density in the $(\Delta, M_2)$ plane. The states populate a compact region whose boundary ABCDEFGHI is the ``Pareto frontier'' of magic and entanglement mentioned in the title. The interesting points and features along this Pareto frontier have been labelled with Latin letters, starting from the origin and moving counterclockwise. 

The main goal of this paper is to describe analytically the Pareto boundary ABCDEFGHI and to identify the corresponding (families of) quantum states along it. This is a familiar task for particle physicists, who often deal with similar restricted compact regions, namely, the kinematically allowed phase space for multi-particle final states \cite{Barr:2011xt,Franceschini:2022vck}. The corresponding kinematic boundaries in many-body phase space result from very special momentum configurations of the particles involved \cite{Cho:2008tj,Matchev:2009fh,Kim:2017awi,Matchev:2019sqa}, and the features along those boundary lines (e.g., sharp kinks) encode useful information about the underlying physics model \cite{Barr:2007hy,Cho:2007dh,Cheng:2008hk,Burns:2008va,Burns:2009zi,Barr:2009jv,Debnath:2016gwz}. Our main objective here is to perform a similar analysis of the ``phase space'' of quantum states, prioritizing analytical descriptions wherever possible, and minimizing reliance on purely numerical methods.

The density distribution depicted in Fig.~\ref{fig:M2_versus_Delta} reveals several interesting boundary features which will be our main object of study:
\begin{itemize}
\item \textbf{The four cornerstones (points A, C, D and I).} The allowed region in the $(\Delta, M_2)$ plane is anchored by four special points, A, C, D and I. Points A and I delineate the range of magic for separable states (with zero entanglement), while points C and D bookend the range of magic for maximally entangled states, i.e., states with $\Delta=1$. 
In particular, point D represents maximally entangled states with maximal magic, while point I represents separable states with maximal magic.
\item \textbf{The Pareto frontier of minimal magic (the ABC boundary).}
A notable feature in Fig.~\ref{fig:M2_versus_Delta} is the existence of a {\it lower bound} on the magic $M_2$ for a given entanglement $\Delta$. In other words, {\it partially entangled states are necessarily magical}, and only separable states or maximally entangled states can have zero magic. In Section~\ref{sec:LowerPareto} we identify the complete set of minimally magical states and derive an analytical expression for the ABC boundary:
\begin{equation}
M_2^{(\text{min})}(\Delta) 
\equiv f_\text{ABC}(\Delta) 
= -\ln \left(\Delta^4-\Delta^2+1\right) 
= \ln \left(\frac{4}{(2\Delta^2-1)^2 + 3}\right),
\label{eq:M2min_delta}
\end{equation}
which is shown with the green line in Fig.~\ref{fig:M2_versus_Delta}.
\item \textbf{The Pareto frontier of maximal magic (the IHGFED boundary).} Similarly, for a given level of entanglement, there also exists {\it an upper bound} on the magic $M_2$, which is given by the boundary IHGFED. We find that this line consists of three separate segments: IHG, GFE, and ED. In Section~\ref{sec:UpperPareto} we identify the corresponding maximally magical states and derive the following analytic expressions for the upper Pareto boundary:
\begin{equation}
M_2^{(\text{max})}(\Delta) = 
\left\{ 
\begin{array}{cl}
f_\text{IHG}(\Delta), & \text{for} \ \Delta\le\Delta_\text{G} \approx 0.63726445;\\[2mm]
f_\text{GFE}(\Delta), & \text{for} \ \Delta_\text{G}\le\Delta\le \Delta_\text{E} =\sqrt{\frac{3}{4}}; \\[2mm]
f_\text{ED}(\Delta),  & \text{for} \ \Delta \ge \Delta_\text{E}.
\end{array}
\right.
\label{eq:M2max_delta}
\end{equation}
where 
\begin{subequations}
\begin{eqnarray}
f_\text{IHG}(\Delta) &=& 
\ln \left(\frac{9}{3\Delta^4 - 2\Delta^3 +4}\right)
= \ln \left(\frac{144}{(12\Delta^2+4\Delta+1)(2\Delta-1)^2 +63}\right),
\label{eq:fIHG}\\ [2mm]
f_\text{GFE}(\Delta) &=& 
\ln \left(\frac{16}{8\Delta^4 - 8\Delta^2 +9}\right)
= \ln \left(\frac{16}{2(2\Delta^2-1)^2+7}\right), \label{eq:fGFE}\\ [2mm]
f_\text{ED}(\Delta) &=& 
\ln \left(\frac{18}{7\Delta^4 - 6\Delta^2 +9}\right)
. \label{eq:fED}
\end{eqnarray}
\label{eq:f_functions}
\end{subequations}
These three lines are depicted in Fig.~\ref{fig:M2_versus_Delta} with the red, yellow and black line, respectively. We emphasize that the expressions in eqs.~(\ref{eq:M2min_delta}-\ref{eq:f_functions}) were obtained analytically and not through a numerical fit.
\item \textbf{The local maxima (points B, H and F).} The Pareto frontiers ABC and IHGFED exhibit local maxima at point B, and at points H and F, respectively. 
\begin{itemize}
\item \textbf{Max-min magic.} Point B represents the maximal value of the minimal magic. From eq.~(\ref{eq:M2min_delta}) it is easy to see that this maximal value is $M_2=\ln\frac{4}{3}=0.287682...$ and is obtained at $\Delta_\text{B}=\frac{1}{\sqrt{2}}=0.70710678...$
\item \textbf{Max-max magic.} Given the practical importance of magic for quantum algorithms, the question of the largest possible value of $M_2$ has been receiving a lot of attention recently.\footnote{For a recent study of the {\em average} value of the magic as a function of entanglement, see \cite{Iannotti:2025lkb}.} For the two-qubit case considered here, a recent paper by Liu, Low and Yin \cite{Liu:2025frx} tightened the previous known upper bound of $\ln \frac{5}{2}=0.916$ to $\ln \frac{16}{7}=0.827$ and showed that this maximum is achieved at two special values of the concurrence: $\Delta = 1/2$ and $\Delta = 1/\sqrt{2}$. Our results for the upper boundary IHGFED are in perfect agreement with those in Ref.~\cite{Liu:2025frx}: eqs.~(\ref{eq:fIHG}) and (\ref{eq:fGFE}) show that the maximal value of $M_2=\ln{\frac{144}{63}}=\ln{\frac{16}{7}}$ is indeed obtained for $\Delta_\text{H}=1/2$ and $\Delta_\text{F}=1/\sqrt{2}$, respectively.
\end{itemize}
\item \textbf{The kinks (points G and E).} Since the upper Pareto boundary IHGFED is piecewise defined through eq.~(\ref{eq:M2max_delta}), there are two special points G and E, where two different branches intersect. At point G, this results in a ``kink'' feature which is familiar to particle physicists from similar phenomena along kinematic phase space boundaries~\cite{Barr:2007hy,Cho:2007dh,Cheng:2008hk,Burns:2008va,Burns:2009zi,Barr:2009jv}. In Section~\ref{sec:UpperPareto}, we show that the $\Delta$-location of point G is given by the second-largest solution to the quartic equation $24 \Delta^4 + 32 \Delta^3 - 72 \Delta^2 + 17 = 0$ (an analytic formula for this solution is available, but is not very illuminating). On the other hand, we find that at point E (located at $\Delta_\text{E}=\sqrt{3/4}$), where the boundaries GFE and ED merge, their slopes are equal:
\begin{equation}
\left(\frac{\partial f_\text{GFE}}{\partial \Delta} \right)_{\text{point E}}   =
\left(\frac{\partial f_\text{ED}}{\partial \Delta} \right)_{\text{point E}} = - \frac{16}{15}\sqrt{\frac{3}{4}}, 
\label{eq:GFE_ED_merger}
\end{equation}
as can be easily verified with eqns.~(\ref{eq:fGFE}) and (\ref{eq:fED}). Therefore, point E is not a true kink point, but rather a {\em point of tangency}. 
\end{itemize}

Another noteworthy feature of the density distribution in Fig.~\ref{fig:M2_versus_Delta} is that the regions in the immediate vicinity of the two Pareto frontiers are relatively sparsely populated.\footnote{Contrast this with the case of phase-space projections in particle physics onto a lower-dimensional manifold, where the resulting number density often peaks (and is sometimes even singular) on the kinematic boundaries. This phenomenon motivated the introduction of a useful class of kinematic variables called singularity variables \cite{Kim:2009si,Rujula:2011qn,DeRujula:2012ns,Agrawal:2013uka,Kim:2019prx,Matchev:2019bon,Park:2020rol,Park:2021lwa}.} 
This implies that randomly generated states are unlikely to approach the extreme values of $M_2$. On the other hand, our analytical results in Sections~\ref{sec:LowerPareto} and \ref{sec:UpperPareto} provide direct access to states which by construction are guaranteed to have extreme magic, i.e., either the largest possible $M_2$ or the lowest possible $M_2$ at a given $\Delta$. 

The paper is organized as follows. In Section~\ref{sec:notation}, we go over our conventions and notation: in Section~\ref{sec:parametrization}, we review the natural analytical parametrization of a two-qubit state introduced by Wharton in \cite{Wharton:2016njz}, while in Sections~\ref{sec:entanglement} and \ref{sec:magic}, we define our quantitative measures of entanglement and magic, respectively. 
Then in Sections~\ref{sec:LowerPareto} and \ref{sec:UpperPareto}, we proceed to study the lower Pareto boundary ABC and the upper Pareto boundary IHGFED, respectively. 
In each case, we identify the relevant set of quantum states, and derive the analytical expressions for the boundary and its features. 
We summarize our results and discuss their implications in Section~\ref{sec:conclusions}.

\section{Conventions and Notation}
\label{sec:notation}

\subsection{Parametrization of two-qubit states}
\label{sec:parametrization}

A general state $\vert\psi\rangle$ of two qubits is a superposition of the four computational basis states
\begin{equation}
\vert \psi \rangle = 
a \vert 00 \rangle 
+b \vert 01 \rangle 
+c \vert 10 \rangle 
+d \vert 11 \rangle \,,
\label{eq:generic_psi_def}
\end{equation}
with some complex coefficients $a$, $b$, $c$ and $d$, normalized as $|a|^2+|b|^2+|c|^2+|d|^2=1$. Eliminating a global phase reduces the 2-qubit parameter space to a six-dimensional manifold~\cite{Kus:2000uot}, for which we find it convenient to use the ``natural'' 6-angle parameterization introduced by Wharton \cite{Wharton:2016njz}:
\begin{subequations}
\label{eq:abcd}
\begin{align}
a=\left[ \cos\frac{\chi}{2}\cos\frac{\theta_1}{2}\cos\frac{\theta_2}{2}e^{i\gamma/2}+\sin\frac{\chi}{2}\sin\frac{\theta_1}{2}\sin\frac{\theta_2}{2}e^{-i\gamma/2} \right] e^{-i(\phi_1+\phi_2)/2}, \\
b=\left[ \cos\frac{\chi}{2}\cos\frac{\theta_1}{2}\sin\frac{\theta_2}{2}e^{i\gamma/2}-\sin\frac{\chi}{2}\sin\frac{\theta_1}{2}\cos\frac{\theta_2}{2}e^{-i\gamma/2} \right] e^{-i(\phi_1-\phi_2)/2},  \\
c=\left[ \cos\frac{\chi}{2}\sin\frac{\theta_1}{2}\cos\frac{\theta_2}{2}e^{i\gamma/2}-\sin\frac{\chi}{2}\cos\frac{\theta_1}{2}\sin\frac{\theta_2}{2}e^{-i\gamma/2} \right] e^{+i(\phi_1-\phi_2)/2},  \\
d=\left[ \cos\frac{\chi}{2}\sin\frac{\theta_1}{2}\sin\frac{\theta_2}{2}e^{i\gamma/2}+\sin\frac{\chi}{2}\cos\frac{\theta_1}{2}\cos\frac{\theta_2}{2}e^{-i\gamma/2} \right] e^{+i(\phi_1+\phi_2)/2}. 
\end{align}
\end{subequations}
The six angles $(\theta_1,\phi_1,\theta_2,\phi_2,\chi,\gamma)$ can be understood as follows \cite{Wharton:2016njz}.
\begin{itemize}
\item The angles $\theta_i$ and $\phi_i$ ($i=1,2$) parametrize two normalized local spinors $\vert\psi_i\rangle$
\begin{subequations}
\label{eq:spinors_def}
\begin{eqnarray}
\vert \psi_1 \rangle &=& e^{-\alpha_1/2}
\left(
\begin{array}{c}
\cos\frac{\theta_1}{2}\, e^{-i\phi_1/2} \\
\sin\frac{\theta_1}{2}\, e^{+i\phi_1/2}
\end{array}
\right)
\equiv 
\left(
\begin{array}{c}
A \\
B
\end{array}
\right), \\ [2mm]
\vert \psi_2 \rangle &=& e^{-\alpha_2/2}
\left(
\begin{array}{c}
\cos\frac{\theta_2}{2}\, e^{-i\phi_2/2} \\
\sin\frac{\theta_2}{2}\, e^{+i\phi_2/2}
\end{array}
\right)
\equiv 
\left(
\begin{array}{c}
C \\
D
\end{array}
\right), 
\end{eqnarray}
\end{subequations}
where $\alpha_1$ and $\alpha_2$ are global phases.
\item The {\em concurrence angle} $\chi$ parametrizes the entanglement between the two qubits since the general state (\ref{eq:generic_psi_def}) can be written as follows
\begin{subequations}
\label{eq:chi_def}
\begin{eqnarray}
a &=& AC\cos\frac{\chi}{2} + B^\ast D^\ast \sin\frac{\chi}{2}, \\ [2mm]
b &=& AD\cos\frac{\chi}{2} - B^\ast C^\ast \sin\frac{\chi}{2}, \\ [2mm]
c &=& BC\cos\frac{\chi}{2} - A^\ast D^\ast \sin\frac{\chi}{2}, \\ [2mm]
d &=& BD\cos\frac{\chi}{2} + A^\ast C^\ast \sin\frac{\chi}{2}.
\end{eqnarray}
\end{subequations}
The {\em concurrence} $\Delta$ (defined below in eq.~(\ref{eq:delta_def})) depends only on the angle $\chi$, which justifies the name ``concurrence angle'' for $\chi$.

\item Note that the parametrization (\ref{eq:chi_def}) depends only on the sum of $\alpha_1$ and $\alpha_2$ and not on their difference. Thus the remaining parameter $\gamma$ is the \emph{recurrence angle} \cite{Wharton:2016njz}, which is simply given by
\begin{equation}
\gamma = \alpha_1+\alpha_2.
\label{eq:gamma_def}
\end{equation}
With the global phase chosen such that the complex concurrence $2(ad-bc)$ is real, one convenient formula for $\gamma$ is
\begin{equation}
\sin\gamma=\frac{2\,\mathrm{Im}(ad+bc)}{\cos\chi\,\sin\theta_1\,\sin\theta_2}.
\end{equation}
It is easy to check that substituting (\ref{eq:spinors_def}) and (\ref{eq:gamma_def}) into (\ref{eq:chi_def}) yields the natural parametrization (\ref{eq:abcd}) advocated in Ref.~\cite{Wharton:2016njz}.
\end{itemize}


\subsection{Measures of entanglement}
\label{sec:entanglement}

There are several useful quantitative measures of entanglement discussed in the literature~\cite{Plenio:2007zz}. Following~\cite{Liu:2025frx}, here we use the {\em concurrence} $\Delta$~\cite{Hill:1997pfa,Wootters:1997id}
\begin{equation}
\Delta \equiv 2 \left| ad - bc\right| = \sin\chi.
\label{eq:delta_def}
\end{equation}
It vanishes for separable states ($\chi=0$) and is maximal and equal to 1 for maximally entangled states ($\chi=\frac{\pi}{2}$). In light of the relation (\ref{eq:delta_def}), the parameters $\Delta$ and $\chi$ can be used interchangeably when it comes to quantifying entanglement. For consistency, throughout the paper we shall stick to using $\Delta$, although our results can be easily rewritten in terms of the concurrence angle $\chi$. 

\subsection{Measures of magic}
\label{sec:magic}

To quantify magic as a quantum resource, we shall follow \cite{Liu:2025frx} and focus on \emph{stabilizer R\'enyi entropies} $M_\alpha$~\cite{Leone:2021rzd} and in particular the order-$2$ case $M_2$~\cite{Oliviero:2022bqm}, which for $n$ qubits is defined as 

\begin{equation}
    M_2(\psi) = -\ln \left( \sum_{P_1,P_2 \in \{I,X,Y,Z\}} \frac{\vert \langle \psi \vert P_1\otimes P_2 \vert \psi \rangle \vert^4}{2^n}  \right)\,.
    \label{eq:M2def}
\end{equation}
By definition, $M_\alpha(\psi)\ge 0$; it vanishes on stabilizer states and increases with non-stabilizer structure, making it a computationally convenient scalar quantifier of magic. The operators $P_1$ and $P_2$ are chosen from the set of operators from the Pauli group: the identity operator $I$ and the Pauli matrices $X\equiv \sigma_x$, $Y\equiv \sigma_y$, and $Z\equiv\sigma_z$.

Note that the sum in eq.~(\ref{eq:M2def}) contains a total of 16 terms involving expectation values of various $P_1\otimes P_2$ operators. It is straightforward to show that these expectation values are given in terms of the 6 angles from the natural parametrization (\ref{eq:abcd}) as follows 
\begin{subequations}
\allowdisplaybreaks
\begin{eqnarray}
\langle \psi \vert I\otimes I \vert \psi \rangle &=& 1, \label{eq:MII}\\ [2pt]
\langle \psi \vert X\otimes I \vert \psi \rangle &=& \cos\chi\, \sin\theta_1\, \cos\phi_1, \label{eq:MXI}\\
\langle \psi \vert Y\otimes I \vert \psi \rangle &=& \cos\chi\, \sin\theta_1\, \sin\phi_1, \label{eq:MYI} \\
\langle \psi \vert Z\otimes I \vert \psi \rangle &=& \cos\chi\, \cos\theta_1, \label{eq:MZI} \\ [2pt]
\langle \psi \vert I\otimes X \vert \psi \rangle &=& \cos\chi\, \sin\theta_2\, \cos\phi_2, \label{eq:MIX} \\
\langle \psi \vert I\otimes Y \vert \psi \rangle &=& \cos\chi\, \sin\theta_2\, \sin\phi_2, \label{eq:MIY}  \\
\langle \psi \vert I\otimes Z \vert \psi \rangle &=& \cos\chi\, \cos\theta_2, \label{eq:MIZ}  \\ [2pt]
\langle \psi \vert X\otimes X \vert \psi \rangle 
&=& \sin\chi\, \cos\theta_1\, \cos\theta_2\, \cos\phi_1\, \cos\phi_2\, \cos\gamma \nonumber \\
&+& \sin\chi\, \cos\theta_1\, \cos\phi_1\, \sin\phi_2\, \sin\gamma \nonumber \\ 
&+& \sin\chi\, \cos\theta_2\, \sin\phi_1\, \cos\phi_2\, \sin\gamma \nonumber \\ 
&-& \sin\chi\, \sin\phi_1\, \sin\phi_2\, \cos\gamma \nonumber \\ 
&+& \sin\theta_1\, \sin\theta_2\, \cos\phi_1\, \cos\phi_2,  \label{eq:MXX}  \\ 
\langle \psi \vert Y\otimes Y \vert \psi \rangle 
&=&  \sin\chi\, \cos\theta_1\, \cos\theta_2\, \sin\phi_1\, \sin\phi_2\, \cos\gamma \nonumber \\
&-& \sin\chi\, \cos\theta_1\, \sin\phi_1\, \cos\phi_2\, \sin\gamma \nonumber \\ 
&-& \sin\chi\, \cos\theta_2\, \cos\phi_1\, \sin\phi_2\, \sin\gamma \nonumber \\ 
&-& \sin\chi\, \cos\phi_1\, \cos\phi_2\, \cos\gamma \nonumber \\ 
&+& \sin\theta_1\, \sin\theta_2\, \sin\phi_1\, \sin\phi_2,  \label{eq:MYY} \\
\langle \psi \vert Z\otimes Z \vert \psi \rangle &=& \sin\chi\, \sin\theta_1\, \sin\theta_2\, \cos\gamma + \cos\theta_1\, \cos\theta_2, \label{eq:MZZ} \\
\langle \psi \vert X\otimes Y \vert \psi \rangle 
&=& \sin\chi\, \cos\theta_1\, \cos\theta_2\, \cos\phi_1\, \sin\phi_2\, \cos\gamma \nonumber \\
&-& \sin\chi\, \cos\theta_1\, \cos\phi_1\, \cos\phi_2\, \sin\gamma \nonumber \\ 
&+& \sin\chi\, \cos\theta_2\, \sin\phi_1\, \sin\phi_2\, \sin\gamma \nonumber \\ 
&+& \sin\chi\, \sin\phi_1\, \cos\phi_2\, \cos\gamma \nonumber \\ 
&+& \sin\theta_1\, \sin\theta_2\, \cos\phi_1\, \sin\phi_2,  \label{eq:MXY}\\
\langle \psi \vert Y\otimes X \vert \psi \rangle  
&=& \sin\chi\, \cos\theta_1\, \cos\theta_2\, \sin\phi_1\, \cos\phi_2\, \cos\gamma \nonumber \\
&+& \sin\chi\, \cos\theta_1\, \sin\phi_1\, \sin\phi_2\, \sin\gamma \nonumber \\ 
&-& \sin\chi\, \cos\theta_2\, \cos\phi_1\, \cos\phi_2\, \sin\gamma \nonumber \\ 
&+& \sin\chi\, \cos\phi_1\, \sin\phi_2\, \cos\gamma \nonumber \\ 
&+& \sin\theta_1\, \sin\theta_2\, \sin\phi_1\, \cos\phi_2, \label{eq:MYX} \\
\langle \psi \vert X\otimes Z \vert \psi \rangle  
&=& - \sin\chi\, \cos\theta_1\, \sin\theta_2\, \cos\phi_1\, \cos\gamma \nonumber \\
&-& \sin\chi\, \sin\theta_2\, \sin\phi_1\, \sin\gamma \nonumber \\ 
&+& \sin\theta_1\, \cos\theta_2\, \cos\phi_1,  \label{eq:MXZ} \\
\langle \psi \vert Z\otimes X \vert \psi \rangle  
&=& - \sin\chi\, \sin\theta_1\, \cos\theta_2\, \cos\phi_2\, \cos\gamma \nonumber \\
&-& \sin\chi\, \sin\theta_1\, \sin\phi_2\, \sin\gamma \nonumber \\ 
&+& \cos\theta_1\, \sin\theta_2\, \cos\phi_2, \label{eq:MZX} \\
\langle \psi \vert Y\otimes Z \vert \psi \rangle 
&=& - \sin\chi\, \cos\theta_1\, \sin\theta_2\, \sin\phi_1\, \cos\gamma \nonumber \\
&+& \sin\chi\, \sin\theta_2\, \cos\phi_1\, \sin\gamma \nonumber \\ 
&+& \sin\theta_1\, \cos\theta_2\, \sin\phi_1,  \label{eq:MYZ} \\
\langle \psi \vert Z\otimes Y \vert \psi \rangle 
&=& - \sin\chi\, \sin\theta_1\, \cos\theta_2\, \sin\phi_2\, \cos\gamma \nonumber \\
&+& \sin\chi\, \sin\theta_1\, \cos\phi_2\, \sin\gamma \nonumber \\ 
&+& \cos\theta_1\, \sin\theta_2\, \sin\phi_2. \label{eq:MZY}  
\end{eqnarray}
\label{eq:MABs}
\end{subequations}
The analytical expressions derived in eqs.~(\ref{eq:MABs}) represent a core contribution of this work. They make it easy to derive and interpret the results detailed in the subsequent two sections.

\section{Lower Pareto Frontier: Minimally Magical Entangled States}
\label{sec:LowerPareto}

\begin{figure}[t]
    \centering
    \includegraphics[width=\textwidth]{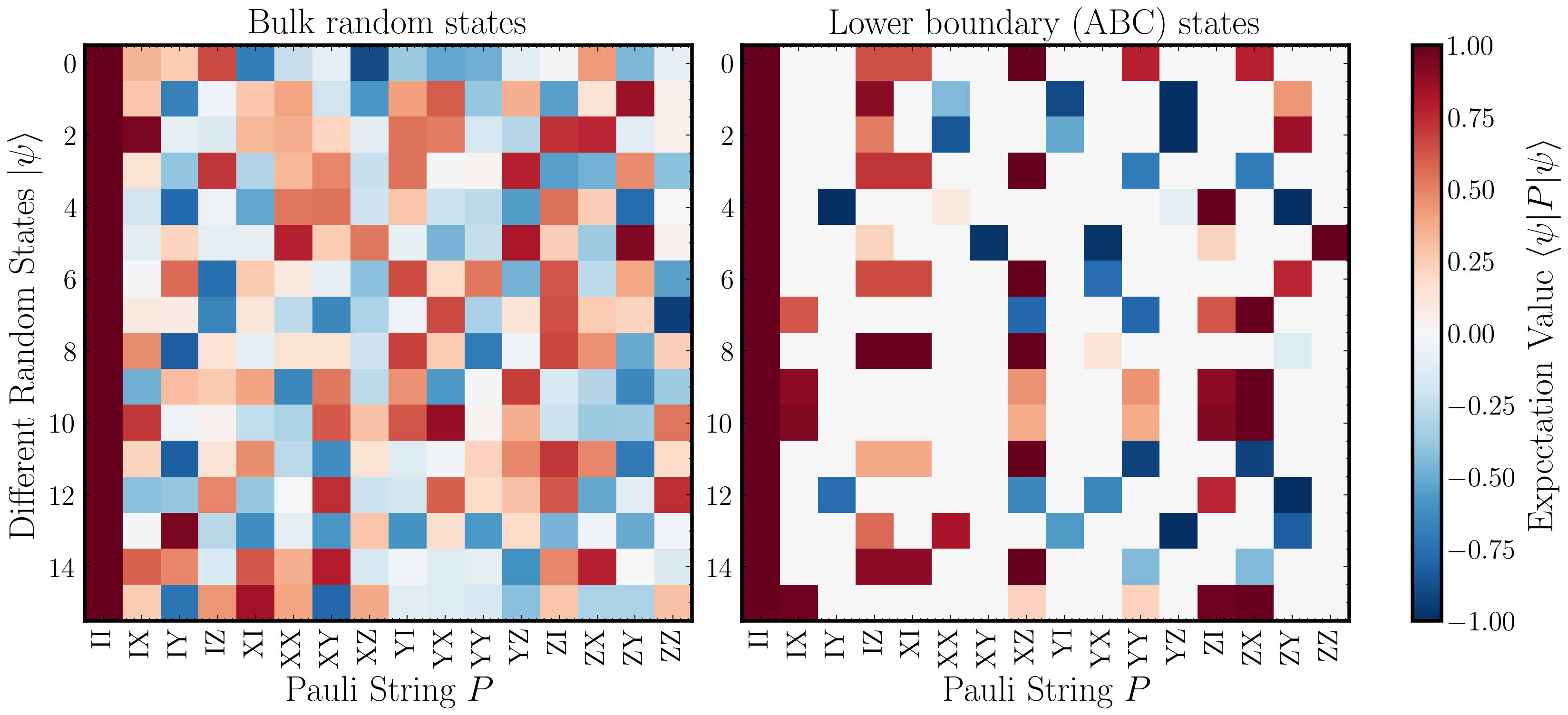}
    \caption{The 16 individual matrix element values (\ref{eq:MABs}) contributing to the sum in (\ref{eq:M2def}) for 16 randomly chosen states in the bulk (left panel) or on the lower Pareto boundary ABC (right panel).}
    \label{fig:bulk}
\end{figure}

For a generic state $\vert \psi\rangle$, all elements in eqs.~(\ref{eq:MABs}) are typically nonzero and range between -1 and 1, as illustrated in the left panel of Figure~\ref{fig:bulk} for 16 randomly chosen states in the bulk. Once raised to the fourth power, all terms from (\ref{eq:MABs}) add coherently, i.e., each element gives a positive contribution to the total magic sum (\ref{eq:M2def}). 
Therefore, for generic (e.g., randomly chosen) states like those in the left panel of Figure~\ref{fig:bulk}, the values of the magic are neither minimal nor maximal, but in between the two extremes, see Figure~\ref{fig:M2_versus_Delta}.

However, on the lower boundary ABC, the situation changes. 
In the right panel of Figure~\ref{fig:bulk}, we show the analogous contribution pattern for 16 randomly chosen states residing {\it on the lower boundary} ABC.
We notice that the pattern is much sparser than the one seen in the left panel.
In particular, for each state, there are exactly 6 non-zero elements from eq.~(\ref{eq:MABs}) contributing to the magic, while the remaining 10 terms are all identically zero. This fact is easy to understand --- since we are trying to minimize a sum of positive-definite contributions, cancellations among terms are impossible and the sum is minimized when certain sets of terms take the lowest possible value, namely 0. 

\begin{figure}[t]
    \centering    
    \includegraphics[width=0.5\linewidth]{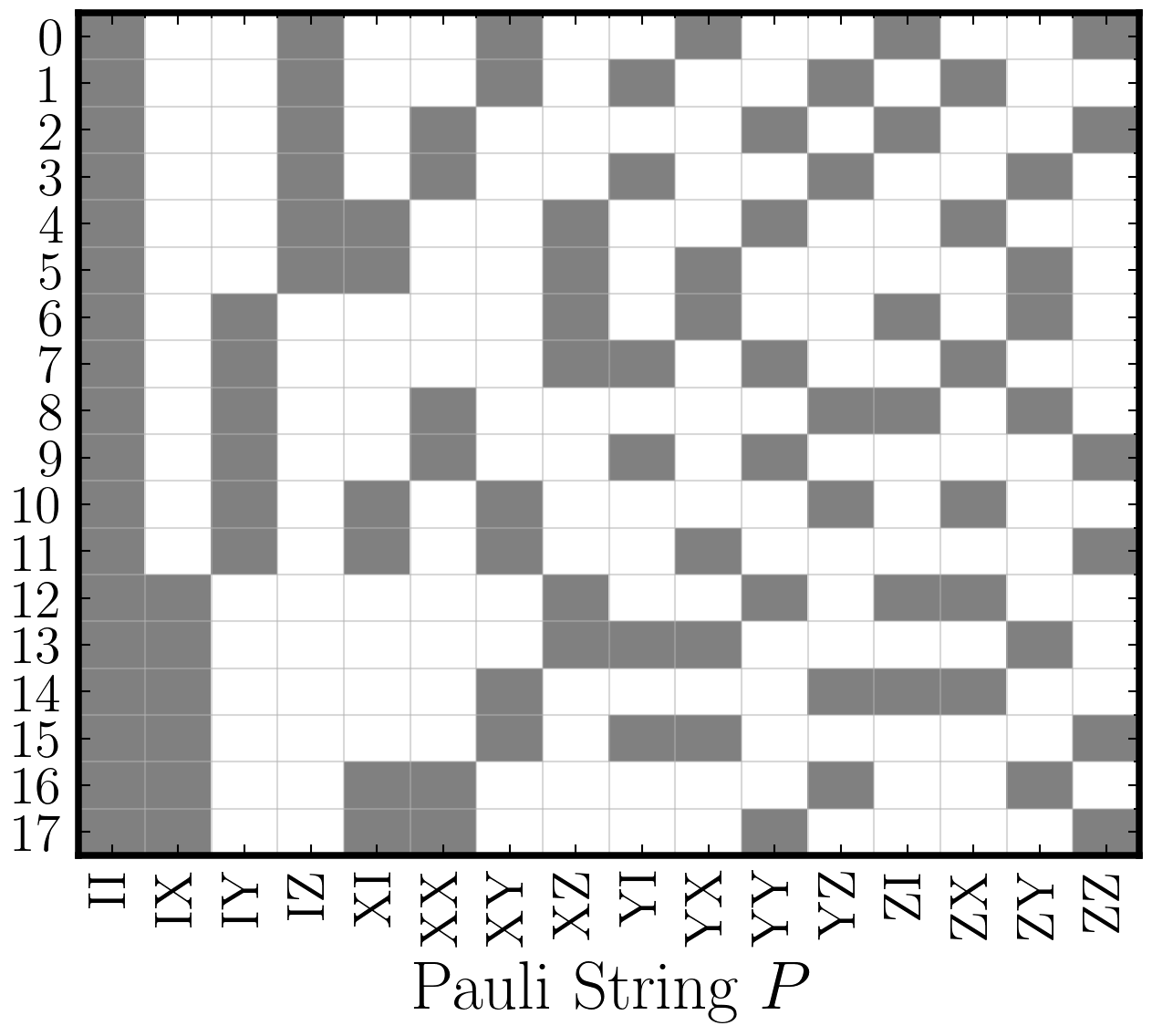} 
    \caption{The 18 patterns of expectation values $\langle\psi\vert P_1\otimes P_2\vert  \psi \rangle$ for the states on the lower Pareto boundary ABC (excluding points A and C). White cells identify the expectation values which are identically zero. The patterns are labelled with python's zero-based indexing  convention, starting from 0.
    }
    \label{fig:lowerpatterns}
\end{figure}

It is interesting to observe that the particular set of ten vanishing terms may vary from one boundary state to another: note that the zero entries in the right panel of Fig.~\ref{fig:bulk} may appear in different locations, for example, states $\{0,3,11,14\}$ have the same pattern of zero elements, while states $\{7, 9, 10, 15\}$ have a different common set of zero terms. Altogether we observe 7 different patterns of zeros in the right panel of Fig.~\ref{fig:bulk}, which brings up the question, how many total such patterns exist. Through an exhaustive parameter search we determined that there are a total of 18 patterns with 10 zeros which we display in Figure~\ref{fig:lowerpatterns} in a binary representation where white (gray) cells indicate zero (non-zero) contributions to the magic sum. We note that in Figure~\ref{fig:lowerpatterns} we do not include the special cases of separable states ($\Delta=0$) and maximally entangled states ($\Delta=1$). For those special states, there are in fact as many as 12 vanishing terms in the magic sum (\ref{eq:M2def}) --- this will become evident very shortly from our discussion below. 

\begin{table}
    \caption{\label{tab:ABC}%
    The parameters of the states with minimal magic. The states are numbered from 0 to 17 in the order in which they appear in Fig.~\ref{fig:lowerpatterns}. The integers $n$ and $m$ count units of $\pi$ and take values 0 or 1, e.g., $\theta_1=n\pi$ means $\theta_1=0$ or $\theta_1=\pi$. }
    \begin{ruledtabular}
    \begin{tabular}{rccccccc}
    \textrm{No}&
    \textrm{Mnemonic}&
    \textrm{Multiplicity}&
    $\theta_1$&
    $\theta_2$&
    $\phi_1$ &
    $\phi_2$ &
    $\gamma$\\
    \colrule
    0  & s s - - - & 8 & $ n\pi$ & $ m\pi$ & any        & any        & $(-1)^n\phi_1+(-1)^m\phi_2\pm\pi/2$   \\
    1  & c s c - - & 8 & $\pi/2$ & $ m\pi$ & $\pm\pi/2$ & any        & $(-1)^m\phi_2+\pi/2\pm\pi/2$  \\
    2  & s s - - - & 8 & $ n\pi$ & $ m\pi$ & any        & any        & $(-1)^n\phi_1+(-1)^m\phi_2+\pi/2\pm\pi/2$   \\
    3  & c s c - - & 8 & $\pi/2$ & $ m\pi$ & $\pm\pi/2$ & any        & $(-1)^m\phi_2\pm\pi/2$   \\
    4  & c s s - - & 8 & $\pi/2$ & $ m\pi$ & $0,\pi$    & any        & $(-1)^m\phi_2+\pi/2\pm\pi/2$  \\
    5  & c s s - - & 8 & $\pi/2$ & $ m\pi$ & $0,\pi$    & any        & $(-1)^m\phi_2\pm\pi/2$  \\
    6  & s c - c - & 8 & $ n\pi$ & $\pi/2$ & any        & $\pm\pi/2$ & $(-1)^n\phi_1+\pi/2\pm\pi/2$  \\
    7  & c c c c c & 8 & $\pi/2$ & $\pi/2$ & $\pm\pi/2$ & $\pm\pi/2$ & $\pm\pi/2$ \\
    8  & s c - c - & 8 & $ n\pi$ & $\pi/2$ & any        & $\pm\pi/2$ & $(-1)^n\phi_1\pm\pi/2$  \\
    9  & c c c c s & 8 & $\pi/2$ & $\pi/2$ & $\pm\pi/2$ & $\pm\pi/2$ & $0,\pi$ \\
    10 & c c s c c & 8 & $\pi/2$ & $\pi/2$ & $0,\pi$    & $\pm\pi/2$ & $\pm\pi/2$ \\
    11 & c c s c s & 8 & $\pi/2$ & $\pi/2$ & $0,\pi$    & $\pm\pi/2$ & $0,\pi$  \\
    12 & s c - s - & 8 & $ n\pi$ & $\pi/2$ & any        & $0,\pi$    & $(-1)^n\phi_1+\pi/2\pm\pi/2$  \\
    13 & c c c s c & 8 & $\pi/2$ & $\pi/2$ & $\pm\pi/2$ & $0,\pi$    & $\pm\pi/2$ \\
    14 & s c - s - & 8 & $ n\pi$ & $\pi/2$ & any        & $0,\pi$    & $(-1)^n\phi_1\pm\pi/2$  \\
    15 & c c c s s & 8 & $\pi/2$ & $\pi/2$ & $\pm\pi/2$ & $0,\pi$    & $0,\pi$ \\
    16 & c c s s c & 8 & $\pi/2$ & $\pi/2$ & $0,\pi$    & $0,\pi$    & $\pm\pi/2$ \\
    17 & c c s s s & 8 & $\pi/2$ & $\pi/2$ & $0,\pi$    & $0,\pi$    & $0,\pi$ \\
    \end{tabular}
    \end{ruledtabular}
\end{table}

Armed with this preliminary knowledge, it is now straightforward to derive {\it exactly} the analytical form of the lower Pareto boundary ABC. To this end, let us choose one of the patterns exhibited in Figure~\ref{fig:lowerpatterns} and try to determine the values of the angles $\theta_1$, $\theta_2$, $\phi_1$, $\phi_2$ and $\gamma$ that will cause the corresponding terms in (\ref{eq:MABs}) to vanish. The resulting values of the angles will be conveniently collected in Table~\ref{tab:ABC}, since they label the different minimal magic states on the boundary ABC.

Let us illustrate the derivation with the very last, 17-th, pattern in Figure~\ref{fig:lowerpatterns}. For brevity, from now on we shall refer to the different expectation values by their Pauli strings, e.g., $XY$ shall refer to $\langle \psi \vert X\otimes Y \vert \psi \rangle$, $IZ$ to $\langle \psi \vert I\otimes Z \vert \psi \rangle$, and so on. The 17-th pattern in Fig.~\ref{fig:lowerpatterns} implies the vanishing of the following terms: $\{IY, IZ, XY, XZ, YI, YX, YZ, ZI, ZX, ZY\}$. Specifically, the vanishing of $ZI$ implies that $\cos\theta_1=0$ in light of eq.~(\ref{eq:MZI}), while the vanishing of $IZ$ leads to $\cos\theta_2=0$ due to eq.~(\ref{eq:MIZ}). Furthermore, once we are assured that $\sin\theta_1$ and $\sin\theta_2$ are non-zero, the vanishing of $YI$ implies $\sin\phi_1=0$ in light of eq.~(\ref{eq:MYI}) and the vanishing of $IY$ leads to $\sin\phi_2=0$ due to eq.~(\ref{eq:MIY}). Finally, the vanishing of $ZY$ implies that $\sin\gamma=0$ (once we take into account that $\cos\theta_1=0$ and $\cos\theta_2=0$). Let us collect what we have learned so far from considering just the five expectation values $\{ZI,IZ,YI,IY,ZY\}$:
\begin{equation}
\cos\theta_1=0, \quad \cos\theta_2=0, \quad 
\sin\phi_1=0, \quad \sin\phi_2=0, \quad 
\sin\gamma=0.
\label{eq:solution_ABC_17}
\end{equation}
It is easy to check that this solution automatically sets the remaining five expectation values on our list, namely $\{XY, YX,XZ,ZX,YZ\}$, to zero as well.
In the second column of Table~\ref{tab:ABC}, we include a useful mnemonic to keep track whether the $\sin$ (s) or the $\cos$ (c) of the corresponding angle is zero for a given solution.
Since the angles are listed in the order $\theta_1$, $\theta_2$, $\phi_1$, $\phi_2$ and $\gamma$, the mnemonic for the solution (\ref{eq:solution_ABC_17}) reads ``ccsss", as shown in the table.
The corresponding values for the angles are listed in the remaining columns of the last row of Table~\ref{tab:ABC}.
Since there are two solutions for each of $\phi_1$, $\phi_2$ and $\gamma$, we find that there are a total of $2^3=8$ states corresponding to pattern 17 in Figure~\ref{fig:lowerpatterns}. 

Once we have taken care of the 10 zero elements in pattern 17, it is now time to look at the remaining six non-vanishing elements which would actually contribute to the magic sum in eq.~(\ref{eq:M2def}). Using the solution (\ref{eq:solution_ABC_17}), it is easy to show that
\begin{subequations}
\begin{eqnarray}
\left| \langle \psi \vert I\otimes I \vert \psi \rangle \right|^4
&= \left| \langle \psi \vert X\otimes X \vert \psi \rangle \right|^4 &= 1, \\
\left| \langle \psi \vert X\otimes I \vert \psi \rangle \right|^4
&= \left| \langle \psi \vert I\otimes X \vert \psi \rangle \right|^4 &= \cos^4\chi = \left(1-\Delta^2\right)^2, \\
\left| \langle \psi \vert Y\otimes Y \vert \psi \rangle \right|^4
&= \left| \langle \psi \vert Z\otimes Z \vert \psi \rangle \right|^4 &= \sin^4\chi = \Delta^4. 
\end{eqnarray}
\label{eq:contributions_ABC}
\end{subequations}
Substituting (\ref{eq:contributions_ABC}) into (\ref{eq:M2def}) and simplifying, we find that the equation for the lower Pareto boundary ABC is simply given by
\begin{equation}
M_2^{(\text{min})}(\Delta) 
\equiv f_\text{ABC}(\Delta) 
= -\ln \left(\Delta^4-\Delta^2+1\right) 
= \ln \left(\frac{4}{(2\Delta^2-1)^2 + 3}\right),
\label{eq:M2min_delta_2}
\end{equation}
which is precisely the result (\ref{eq:M2min_delta}) we advertised in the introduction. It is clear that this function is maximized at point B located at
\begin{equation}
\Delta_\text{B}\equiv \text{argmax}_\Delta \left(
f_\text{ABC}(\Delta)  \right)
=\frac{1}{\sqrt{2}}
\end{equation}
and therefore the maximum value of $M_2$ along the ABC boundary is given by
\begin{equation}
M_2^{(\text{min})}(\Delta_\text{B})  
= \ln\left(\frac{4}{3}\right) = 0.28768207245...
\end{equation}
Another useful cross-check is that for both separable states ($\Delta=0$) and for maximally entangled states ($\Delta=1$), eq.~(\ref{eq:M2min_delta_2}) gives $M_2=\ln(4/4)=0$, as expected. Note another nice numerical result at $\Delta=0.5$:
\begin{equation}
M_2^{(\text{min})}\left(\Delta=\frac{1}{2}\right)  
=  \ln\left(\frac{16}{13}\right) = 0.20763936...
\end{equation}

This completes the discussion of the 17-th pattern in Figure~\ref{fig:lowerpatterns}. One can now repeat the same analysis for each of the remaining 17 patterns (labelled from 0 to 16) in Figure~\ref{fig:lowerpatterns}. In each case, requiring the corresponding pattern of zeros leads to a solution for the angles $\theta_1$, $\theta_2$, $\phi_1$, $\phi_2$ and $\gamma$ which is shown in Table~\ref{tab:ABC}. With those solutions, the six remaining non-zero terms are as follows: there are two terms which are constant and equal to 1, there are two terms equal to $\cos^4\chi$, and there are two terms equal to $\sin^4\chi$. While the specific Pauli strings that correspond to these 6 terms vary from one pattern to another, the sum of those 6 terms always results in the formula given by eq.~(\ref{eq:M2min_delta_2}). This means that {\it all} 144 states listed in Table~\ref{tab:ABC} reside on the same ABC boundary. Note that all these 144 states are different, as measured by their pairwise overlap (fidelity) which is required to be less than 1. 

We should also comment on an interesting feature in Table~\ref{tab:ABC}, namely, for some patterns, we find an apparent continuous family of solutions.
However, the apparent extra degree of freedom cancels out and the multiplicity for those cases is still 8.
To illustrate this, consider one of the two most complex cases in Figure~\ref{fig:lowerpatterns} and Table~\ref{tab:ABC}, the 0-th pattern.
The vanishing terms are $\{IX, IY, XI, XX, XZ, YI, YY, YZ, ZX, ZY\}$.
The solution is obtained as follows: the vanishing of $XI$ and $YI$ implies $\sin\theta_1=0$, while the vanishing of $IX$ and $IY$ leads to $\sin\theta_2=0$.
These two constraints automatically lead to the vanishing of $XZ$, $ZX$, $YZ$ and $ZY$ as well, so at this point the only two remaining terms that need to be set to zero are $XX$ and $YY$.
Let us suppose we choose the case of $\theta_1=\theta_2=0$, i.e., $n=m=0$ in the first row of Table~\ref{tab:ABC}. 
It is easy to show that $XX$ and $YY$ become
\begin{subequations}
\begin{eqnarray}
\langle \psi \vert X\otimes X \vert \psi \rangle 
&=& +\cos(\phi_1+\phi_2-\gamma), \\
\langle \psi \vert Y\otimes Y \vert \psi \rangle 
&=& - \cos(\phi_1+\phi_2-\gamma).
\end{eqnarray}
\end{subequations}
They will both vanish if 
\begin{equation}
\cos(\phi_1+\phi_2-\gamma) = 0 \quad \implies \quad 
\phi_1+\phi_2-\gamma = \pm \frac{\pi}{2}
\quad \implies \quad 
\gamma = \phi_1+\phi_2 \pm \frac{\pi}{2},
\end{equation}
which is the result listed in the first row of Table~\ref{tab:ABC} (for the case of $n=m=0$). The corresponding minimal magic state is 
\begin{equation}
\vert \psi \rangle = 
\cos\frac{\chi}{2}\ e^{\pm i \pi/4}\,
\vert 00 \rangle 
+ \sin\frac{\chi}{2}\ e^{\mp i \pi/4}\,
\vert 11 \rangle,
\end{equation}
where it is clear that any dependence on $\phi_1$ and $\phi_2$ has disappeared.

In conclusion, we reiterate the main result of this section: the existence of the lower Pareto boundary ABC. This boundary implies that partially entangled two-qubit states are guaranteed to be magical, with a minimum value for $M_2$ given by eq.~(\ref{eq:M2min_delta_2}), i.e.
\begin{equation}
M_2(\psi) \ge M_2^{(\text{min})}(\Delta), \quad \forall |\psi\rangle \text{ s.t. } \Delta(\psi) \in (0,1).
\end{equation}

\section{Upper Pareto Frontier: Maximally Magical Entangled States}
\label{sec:UpperPareto}

In this section we analyze the upper Pareto frontier, i.e., the boundary of {\it maximal} posible magic $M_2$ for a given concurrence $\Delta$. As can be seen in Figure~\ref{fig:M2_versus_Delta}, this boundary is piecewise defined and is comprised of three separate segments: a left branch IHG (at small to moderate $\Delta$, up to $\Delta_\text{G}$), a middle branch GFE (for $\Delta$ between $\Delta_\text{G}$ and $\Delta_\text{E}$), and a right branch ED (at the largest values of $\Delta$ above $\Delta_\text{E}$). We now discuss each branch in turn. In each case, the analysis will follow the blueprint from the previous section --- inspect the contribution pattern to the magic sum (\ref{eq:M2def}), find the angles $\theta_1$, $\theta_2$, $\phi_1$, $\phi_2$ and $\gamma$ that would create such a pattern, then derive $M_2$ analytically as a function of the concurrence angle $\chi$.

\subsection{The middle branch GFE}
\label{sec:GFE}

\begin{figure}[t]
    \includegraphics[width=0.5\textwidth]{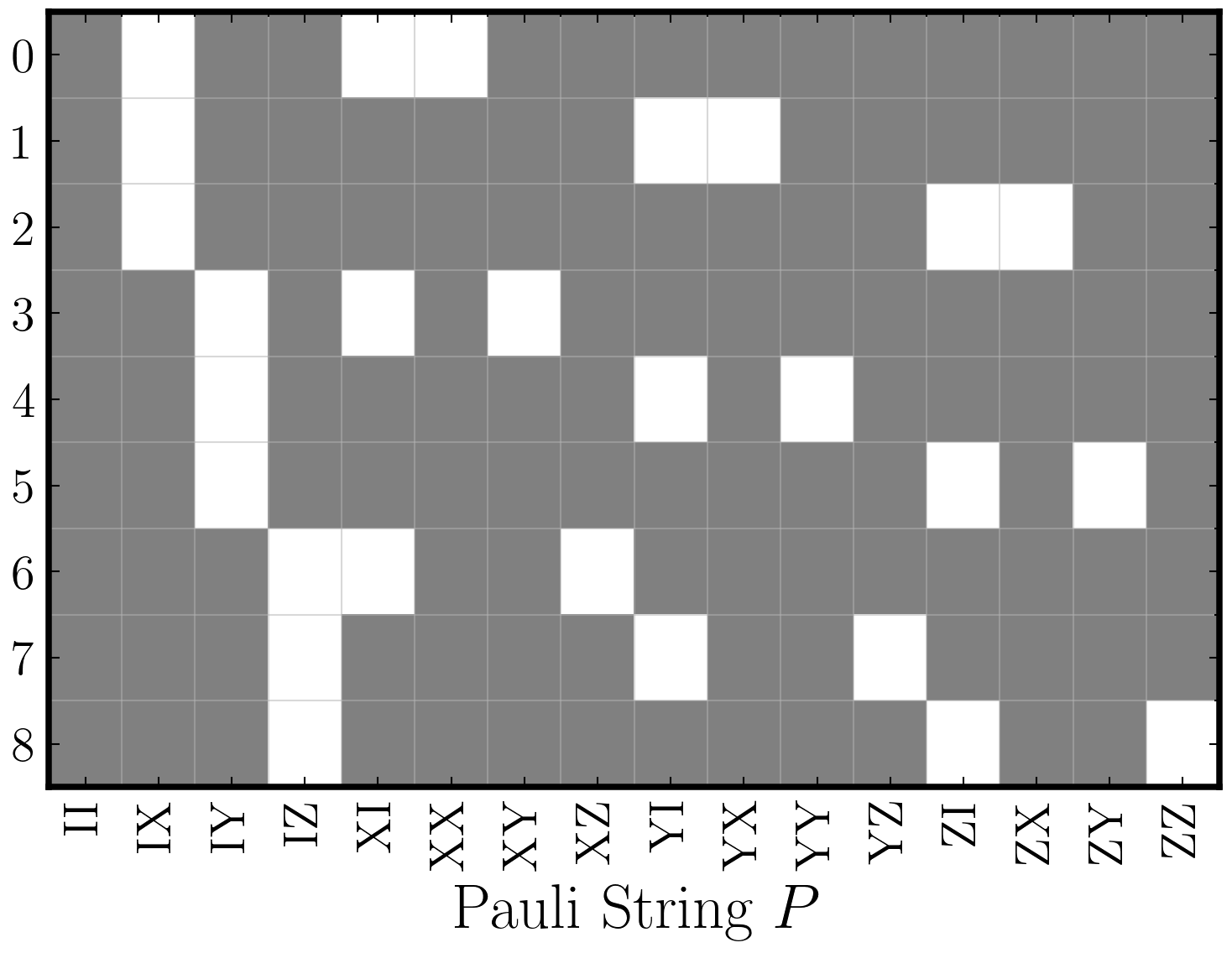}
    \caption{\label{fig:upper_pareto_middle} The 9 patterns for any state on the GFE boundary, including point F. }
    
    \begin{ruledtabular}
    \begin{tabular}{rccccccc}
    \textrm{No}&
    \textrm{Code}&
    \textrm{Multiplicity}&
    $\theta_1$&
    $\theta_2$&
    $\phi_1$ &
    $\phi_2$ &
    $\gamma$\\
    \colrule
    0 & - - c c c & 32 & $\frac{\pi}{4}, \frac{3\pi}{4}$ & $\frac{\pi}{4}, \frac{3\pi}{4}$ & $\frac{\pi}{2}, \frac{3\pi}{2}$ & $\frac{\pi}{2}, \frac{3\pi}{2}$ & $\frac{\pi}{2}, \frac{3\pi}{2}$ \\
    1 & - - s c c & 32 & $\frac{\pi}{4}, \frac{3\pi}{4}$ & $\frac{\pi}{4}, \frac{3\pi}{4}$ & $0, \pi$ & $\frac{\pi}{2}, \frac{3\pi}{2}$ & $\frac{\pi}{2}, \frac{3\pi}{2}$ \\
    2 & c - - c s & 32 & $\frac{\pi}{2}$ & $\frac{\pi}{4}, \frac{3\pi}{4}$ & $\pm\frac{\pi}{4}, \pm\frac{3\pi}{4}$ & $\frac{\pi}{2}, \frac{3\pi}{2}$ & $0,\pi$   \\
    3 & - - c s c & 32 & $\frac{\pi}{4}, \frac{3\pi}{4}$ & $\frac{\pi}{4}, \frac{3\pi}{4}$ & $\frac{\pi}{2}, \frac{3\pi}{2}$ & $0, \pi$ & $\frac{\pi}{2}, \frac{3\pi}{2}$   \\
    4 & - - s s c & 32 & $\frac{\pi}{4}, \frac{3\pi}{4}$ & $\frac{\pi}{4}, \frac{3\pi}{4}$ & $0, \pi$ & $0, \pi$ &  $\frac{\pi}{2}, \frac{3\pi}{2}$ \\
    5 & c - - s s & 32 & $\frac{\pi}{2}$ & $\frac{\pi}{4}, \frac{3\pi}{4}$ & $\pm\frac{\pi}{4}, \pm\frac{3\pi}{4}$ & $0, \pi$ & $0, \pi$  \\
    6 & - c c - s & 32 & $\frac{\pi}{4}, \frac{3\pi}{4}$ & $\frac{\pi}{2}$ & $\frac{\pi}{2}, \frac{3\pi}{2}$ & $\pm\frac{\pi}{4}, \pm\frac{3\pi}{4}$ &  $0, \pi$ \\
    7 & - c s - s & 32 & $\frac{\pi}{4}, \frac{3\pi}{4}$ & $\frac{\pi}{2}$ & $0, \pi$ & $\pm\frac{\pi}{4}, \pm\frac{3\pi}{4}$ & $0,\pi$ \\
    8 & c c - - c & 32 & $\frac{\pi}{2}$ & $\frac{\pi}{2}$ & $\pm\frac{\pi}{4}, \pm\frac{3\pi}{4}$ & $\pm\frac{\pi}{4}, \pm\frac{3\pi}{4}$ & $\frac{\pi}{2}, \frac{3\pi}{2}$ \\
    \end{tabular}
    \end{ruledtabular}
    \captionof{table}{\label{tab:GFE}%
    The parameters of the states with maximal magic on the GFE boundary. The states are numbered from 0 to 8 in the order in which they appear in Fig.~\ref{fig:upper_pareto_middle}.}
\end{figure}

We begin with the middle branch GFE (the orange line in Fig.~\ref{fig:M2_versus_Delta}).
In this case, we find that all states on the GFE boundary fall into one of the nine patterns exhibited in Figure~\ref{fig:upper_pareto_middle}.
Each pattern has exactly three zeros and 13 non-zero terms.
Furthermore, the 12 non-zero terms (other than the trivial $II$) fall into three groups of four equal terms (in absolute value) within each group.
The corresponding solutions for the angles $\theta_1$, $\theta_2$, $\phi_1$, $\phi_2$ and $\gamma$ are given in Table~\ref{tab:GFE}.
The angles $\theta_1$, $\theta_2$, $\phi_1$ and $\phi_2$ are found to be multiples of $\pi/4$, while the angle $\gamma$ is given in multiples of $\pi/2$.
Once we know these angles, we can compute the contribution of each term to the magic sum in (\ref{eq:M2def}).
For illustration, we list the expectation values for the case of the 0-th pattern from Figure~\ref{fig:upper_pareto_middle} and  Table~\ref{tab:GFE}:
\begin{subequations}
\begin{eqnarray}
\left| \langle \psi \vert I\otimes I \vert \psi \rangle \right|^4
&=& 1, \\[2mm]
  \left| \langle \psi \vert Y\otimes Y \vert \psi \rangle \right|^4
= \left| \langle \psi \vert Z\otimes Z \vert \psi \rangle \right|^4 
= \left| \langle \psi \vert Y\otimes Z \vert \psi \rangle \right|^4 
= \left| \langle \psi \vert Z\otimes Y \vert \psi \rangle \right|^4 
&=& \frac{1}{16}, \\[2mm]
  \left| \langle \psi \vert Y\otimes I \vert \psi \rangle \right|^4
= \left| \langle \psi \vert Z\otimes I \vert \psi \rangle \right|^4 
= \left| \langle \psi \vert I\otimes Y \vert \psi \rangle \right|^4 
= \left| \langle \psi \vert I\otimes Z \vert \psi \rangle \right|^4 
&=& \frac{1}{4}\cos^4\chi, \\[2mm]
  \left| \langle \psi \vert X\otimes Y \vert \psi \rangle \right|^4
= \left| \langle \psi \vert Y\otimes X \vert \psi \rangle \right|^4 
= \left| \langle \psi \vert X\otimes Z \vert \psi \rangle \right|^4 
= \left| \langle \psi \vert Z\otimes X \vert \psi \rangle \right|^4 
&=& \frac{1}{4}\sin^4\chi.~~~~~
\end{eqnarray}
\label{eq:contributions_GFE}
\end{subequations}
The remaining three expectation values $IX$, $XI$ and $XX$ are identically zero, as shown in Figure~\ref{fig:upper_pareto_middle}. Substituting (\ref{eq:contributions_GFE}) into (\ref{eq:M2def}) gives
\begin{equation}
M_2^{(\text{max})}(\chi) 
= -\ln \left( \frac{1+\frac{1}{4}+\cos^4\chi+\sin^4\chi}{4} \right) 
= \ln \left(\frac{16}{5+4\cos^4\chi+4\sin^4\chi}\right),
\label{eq:M2max_GFE_chi}    
\end{equation}
and trading $\sin\chi$ for $\Delta$, we finally obtain 
\begin{equation}
M_2^{(\text{max})}(\Delta)
\equiv f_\text{GFE}(\Delta) =
\ln \left(\frac{16}{8\Delta^4 - 8\Delta^2 +9}\right)
= \ln \left(\frac{16}{2(2\Delta^2-1)^2+7}\right), \label{eq:M2max_GFE_Delta} 
\end{equation}
which is the result quoted in eq.~(\ref{eq:fGFE}) of the introduction. 
The maximum at point F is achieved at 
\begin{equation}
\Delta_\text{F}\equiv \text{argmax}_\Delta \left(
f_\text{GFE}(\Delta) \right)
=\frac{1}{\sqrt{2}}    
\end{equation}
and the maximal value of the magic $M_2$ along the middle branch GFE is equal to
\begin{equation}
M_2^{(\text{max})}(\Delta_\text{F}) 
= f_\text{GFE}(\Delta_\text{F}) 
= \ln \left(\frac{16}{7}\right)= 0.82667857318...
\label{eq:pointF_maximum}
\end{equation}
confirming the numerically derived result from \cite{Liu:2025frx}.

We note that Table~\ref{tab:GFE} contains a total of 288 unique states, 32 for each pattern. The states are explicitly labelled by the values of the angles $\theta_1$, $\theta_2$, $\phi_1$, $\phi_2$ and $\gamma$, and are guaranteed to give the largest possible value of $M_2$ for any given $\Delta$ chosen within the domain of the middle branch GFE (between points G and E).

\subsection{The left branch IHG}
\label{sec:IHG}

\begin{figure}[t]
    \includegraphics[width=0.6\textwidth]{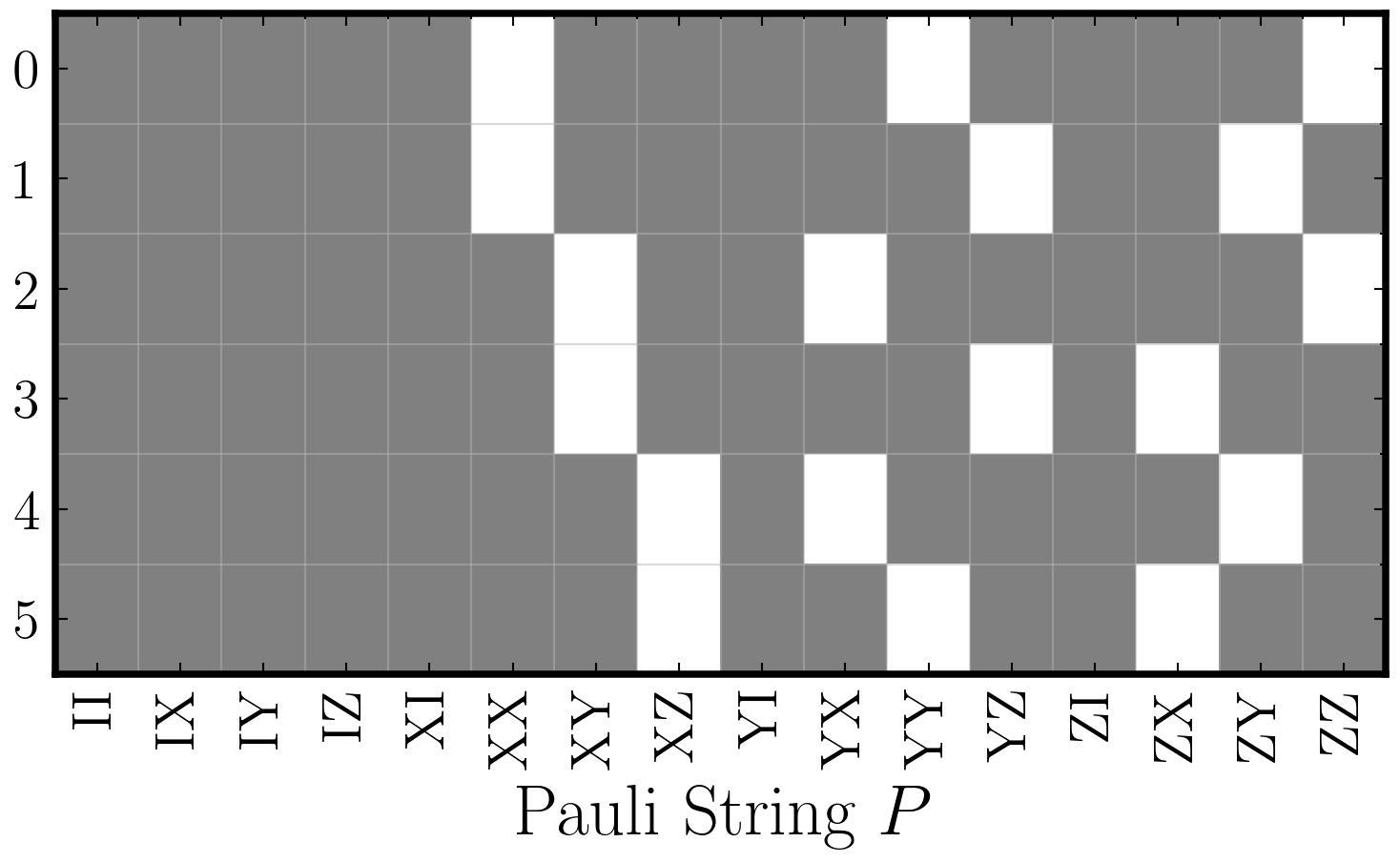}
    \caption{\label{fig:upper_pareto_left} The six patterns for the states at point H on the IHG boundary.  }
    
    \begin{ruledtabular}
    \begin{tabular}{rccccccc}
    \textrm{No}&
    \textrm{Multiplicity}&
    $\theta_1$&
    $\theta_2$&
    $\phi_1$ &
    $\phi_2$ &
    $\gamma$\\
    \colrule
    0 & 48 & $\alpha$ & $\alpha$ & $\pm\frac{\pi}{4}, \pm\frac{3\pi}{4}$  & $\pm\frac{\pi}{4}, \pm\frac{3\pi}{4}$  & $\frac{\pi}{3}, \pi, \frac{5\pi}{3}$ \\
    1 & 48 & $\pi-\alpha$ & $\pi-\alpha$ & $\pm\frac{\pi}{4}, \pm\frac{3\pi}{4}$  & $\pm\frac{\pi}{4}, \pm\frac{3\pi}{4}$  & $\frac{\pi}{3}, \pi, \frac{5\pi}{3}$ \\
    2 & 48 & $\alpha$ & $\pi-\alpha$ & $\pm\frac{\pi}{4}, \pm\frac{3\pi}{4}$  & $\pm\frac{\pi}{4}, \pm\frac{3\pi}{4}$  & $0, \frac{2\pi}{3}, \frac{4\pi}{3}$ \\
    3 & 48 & $\pi-\alpha$ & $\alpha$ & $\pm\frac{\pi}{4}, \pm\frac{3\pi}{4}$  & $\pm\frac{\pi}{4}, \pm\frac{3\pi}{4}$  & $0, \frac{2\pi}{3}, \frac{4\pi}{3}$ 
    \end{tabular}
    \end{ruledtabular}
    \captionof{table}{\label{tab:IHG}%
    The angle parameters of the states with maximal magic on the IHG boundary. $\theta_1$ and $\theta_2$ are defined in terms of $\alpha\equiv \arccos(1/\sqrt{3})=0.9553166181245092...$. }
\end{figure}

We continue with the discussion of the left branch IHG (the red line in Figure~\ref{fig:M2_versus_Delta}). 
In this case we find that, in contrast to the case of the middle branch from the previous subsection, all 16 terms generically contribute to the magic sum (\ref{eq:M2def}).
Besides the trivial $II$ case, the remaining 15 terms fall into three groups of 6, 6 and 3, respectively, where the terms within each group are equal in absolute value.
To organize our presentation in analogy to the other subsections, we shall focus on point H, where the three terms in the last group happen to vanish, which results in patterns with three zeros as before.
We then find that there are 6 different patterns that are depicted in Figure~\ref{fig:upper_pareto_left}.
The corresponding 192 states are listed in Table~\ref{tab:IHG}. Note that $\theta_1$ and $\theta_2$ are always such that $\sin\theta_1=\sin\theta_2=\sqrt{2}/\sqrt{3}$ and are therefore related to the angle $\alpha$ defined as $\alpha\equiv \arccos(1/\sqrt{3})\approx0.9553$.
Table~\ref{tab:IHG} utilizes a compact notation spanning only four rows.
The explicit mapping of these 192 states to the six patterns in Figure~\ref{fig:upper_pareto_left} is provided in Tables~\ref{tab:IHG14} and \ref{tab:IHG56} of Appendix~\ref{app:catalog}. 

Given the values of the angles in Table~\ref{tab:IHG}, we can proceed with the derivation of the IHG boundary. For illustration, we choose the very first case in Table~\ref{tab:IHG}, $\theta_1=\theta_2=\alpha$, $\phi_1=\phi_2=+\pi/4$ and $\gamma=\pi/3$. The 16 expectation values in this case are
\begin{subequations}
\begin{eqnarray}
\left| \langle \psi \vert I\otimes I \vert \psi \rangle \right|^4
&=& 1, \\[2mm]
  \left| \langle \psi \vert X\otimes I \vert \psi \rangle \right|^4
= \left| \langle \psi \vert Y\otimes I \vert \psi \rangle \right|^4 
= \left| \langle \psi \vert Z\otimes I \vert \psi \rangle \right|^4 
&=& \nonumber \\ [2mm]
= \left| \langle \psi \vert I\otimes X \vert \psi \rangle \right|^4 
= \left| \langle \psi \vert I\otimes Y \vert \psi \rangle \right|^4 
= \left| \langle \psi \vert I\otimes Z \vert \psi \rangle \right|^4 
&=& \frac{1}{9} \cos^4\chi, \\[2mm]
  \left| \langle \psi \vert X\otimes X \vert \psi \rangle \right|^4 
= \left| \langle \psi \vert Z\otimes Z \vert \psi \rangle \right|^4 
= \left| \langle \psi \vert X\otimes Y \vert \psi \rangle \right|^4 
&=& \nonumber \\ [2mm]  
= \left| \langle \psi \vert Y\otimes X \vert \psi \rangle \right|^4 
= \left| \langle \psi \vert Y\otimes Z \vert \psi \rangle \right|^4 
= \left| \langle \psi \vert Z\otimes Y \vert \psi \rangle \right|^4 
&=& \frac{1}{81}\,(\sin\chi+1)^4, \\[2mm]
  \left| \langle \psi \vert Y\otimes Y \vert \psi \rangle \right|^4
= \left| \langle \psi \vert X\otimes Z \vert \psi \rangle \right|^4 
= \left| \langle \psi \vert Z\otimes X \vert \psi \rangle \right|^4 
&=& \frac{1}{81}\,(2\sin\chi-1)^4. \label{eq:terms_IHG_4}
\end{eqnarray}
\label{eq:contributions_IHG}
\end{subequations}
Since point H happens to be located at $\Delta_\text{H}=\sin\chi_H=1/2$, the terms in (\ref{eq:terms_IHG_4}) vanish at point H, which results in the three-zero pattern number 5 seen in Figure~\ref{fig:upper_pareto_left} (see also case 5a in Table~\ref{tab:IHG56}). However, elsewhere along the IHG boundary (away from point H), the terms in (\ref{eq:terms_IHG_4}) are in general non-zero. 

Now substituting (\ref{eq:contributions_IHG}) into (\ref{eq:M2def}), we get
\begin{equation}
M_2^{(\text{max})}(\chi) 
= -\ln \left( \frac{1+\frac{2}{3}\cos^4\chi+\frac{6}{81}(\sin\chi+1)^4 + \frac{3}{81}(2\sin\chi-1)^4}{4} 
\right) 
\label{eq:M2max_IHG_chi}    
\end{equation}
and upon trading $\sin\chi$ for $\Delta$ and simplifying, we get
\begin{subequations}
\begin{eqnarray}
M_2^{(\text{max})}(\Delta)
\equiv f_\text{IHG}(\Delta)
&=& \ln \left(\frac{9}{3\Delta^4 - 2\Delta^3 +4}\right)
\\ [2mm]
&=& \ln \left(\frac{144}{(12\Delta^2+4\Delta+1)(2\Delta-1)^2 +63}\right),
\end{eqnarray}
\label{eq:M2max_IHG_Delta} 
\end{subequations}
which is the result quoted in eq.~(\ref{eq:fIHG}) from the introduction. It is obvious that the maximum at point H is achieved for
\begin{equation}
\Delta_\text{H}\equiv \text{argmax}_\Delta \left(
f_\text{IHG}(\Delta) \right)
=\frac{1}{2}    
\end{equation}
and the maximal value of the magic at that point is equal to 
\begin{equation}
M_2^{(\text{max})}(\Delta_\text{H}) 
= f_\text{IHG}(\Delta_\text{H}) 
= \ln \left(\frac{16}{7}\right)= 0.82667857318...
\end{equation}
This happens to be the same value as that in eq.~(\ref{eq:pointF_maximum}), a fact which was also pointed out in Ref.~\cite{Liu:2025frx}.



Now that we have the analytical expressions for the IHG and GFE branches, we can find the location of their intersection point G as a solution to the following equation
\begin{equation}
f_\text{IHG} (\Delta) = f_\text{GFE} (\Delta)
\end{equation}
which leads to the quartic equation
\begin{equation}
24 \Delta^4 + 32 \Delta^3 - 72 \Delta^2 + 17 = 0.
\label{eq:delta_G_equation}
\end{equation}
This equation has four real solutions, two of which are of interest to us as they are located in the relevant region of $\Delta$ between 0 and 1.
Point G is located at\footnote{While eq.~(\ref{eq:delta_G_equation}) admits exact analytical solutions, the resulting expressions are unwieldy and offer little physical insight. 
We therefore report only the numerical values.} $\Delta_\text{G}\approx 0.63726445$.
Note that the extensions of the IHG and GFE boundaries (the red dashed and orange dashed lines in Figure~\ref{fig:M2_versus_Delta}) intersect once more at $\Delta\approx 0.977411$, but there the maximal magic is given by the ED boundary to be discussed in the next subsection.

On the left end of the IHG boundary we find point I, which represents separable states with maximal magic. Since $M_2$ is additive for separable states, the magic of the two-qubit states at point I is equal to twice the maximal magic for a single qubit state, which is $\ln\frac{3}{2}$. 
Therefore, we expect the value of $M_2$ at point I to be $\ln\frac{3}{2}+\ln\frac{3}{2}=\ln\frac{9}{4}$, which is precisely the answer we get from eq.~(\ref{eq:M2max_IHG_Delta}) for $\Delta=0$.

\subsection{The right branch ED}
\label{sec:ED}

\begin{figure}[t]
    \centering
    \includegraphics[width=0.5\linewidth]{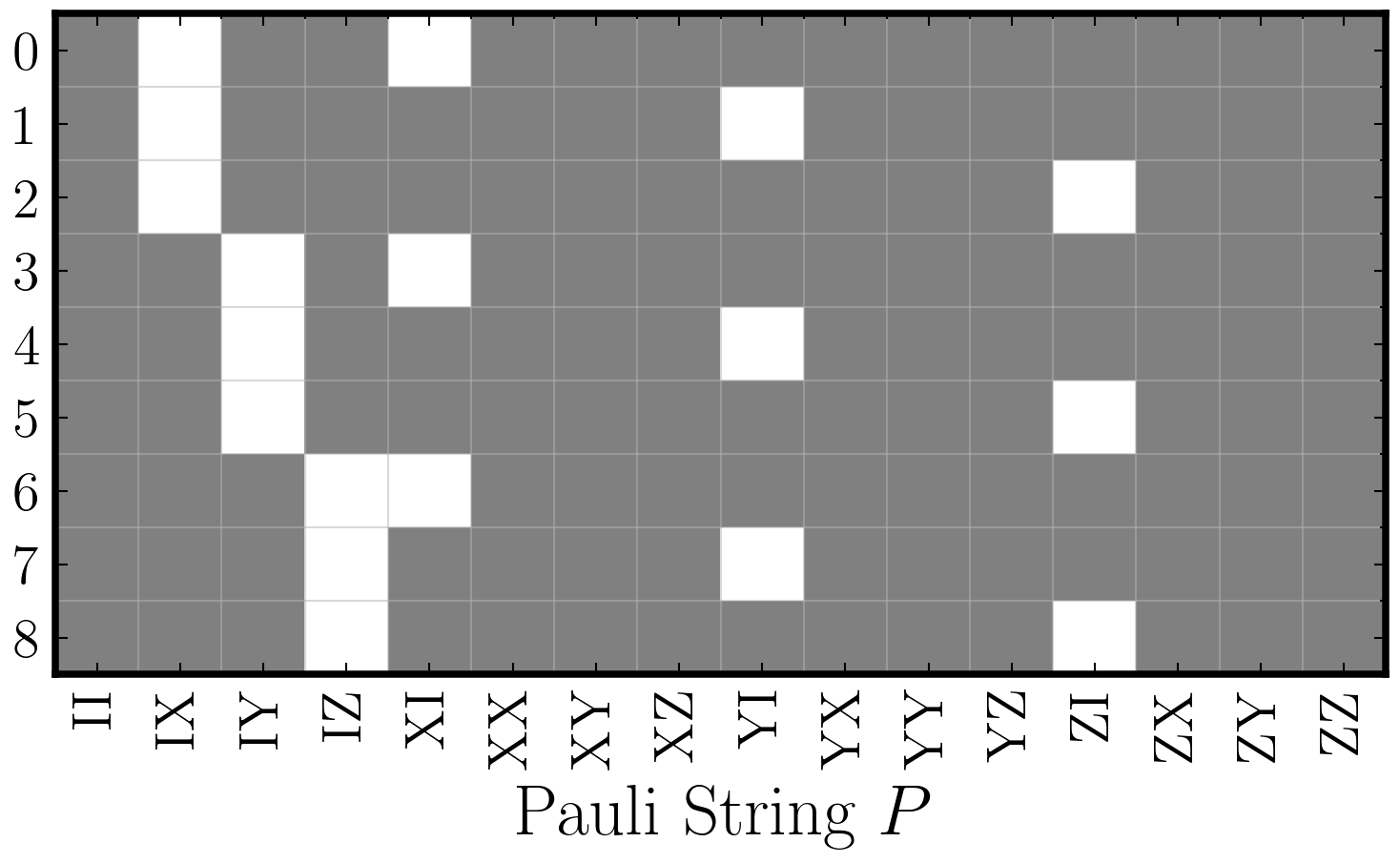}
    \caption{The 9 patterns for states on the ED boundary. }
    \label{fig:upper_pareto_right}
    
    \begin{ruledtabular}
    \begin{tabular}{rcccccc}
    \textrm{No}&
    \textrm{Multiplicity}&
    $\theta_1$&
    $\theta_2$&
    $\phi_1$ &
    $\phi_2$ &
    $\gamma$\\
    \colrule
    0 & 64 & $\frac{\pi}{4}, \frac{3\pi}{4}$ & $\frac{\pi}{4}, \frac{3\pi}{4}$ & $\frac{\pi}{2}, \frac{3\pi}{2}$ & $\frac{\pi}{2}, \frac{3\pi}{2}$ & $\frac{\pi}{2}\pm\delta, \frac{3\pi}{2}\pm\delta$ \\
    1 & 64 & $\frac{\pi}{4}, \frac{3\pi}{4}$ & $\frac{\pi}{4}, \frac{3\pi}{4}$ & $0, \pi$ & $\frac{\pi}{2}, \frac{3\pi}{2}$ & $\frac{\pi}{2}\pm\delta, \frac{3\pi}{2}\pm\delta$ \\
    2 & 64 & $\frac{\pi}{2}$ & $\frac{\pi}{4}, \frac{3\pi}{4}$ & $\pm\frac{\pi}{4}, \pm\frac{3\pi}{4}$ & $\frac{\pi}{2}, \frac{3\pi}{2}$ & $0\pm\delta,\pi\pm\delta$   \\
    3 & 64 & $\frac{\pi}{4}, \frac{3\pi}{4}$ & $\frac{\pi}{4}, \frac{3\pi}{4}$ & $\frac{\pi}{2}, \frac{3\pi}{2}$ & $0, \pi$ & $\frac{\pi}{2}\pm\delta, \frac{3\pi}{2}\pm\delta$   \\
    4 & 64 & $\frac{\pi}{4}, \frac{3\pi}{4}$ & $\frac{\pi}{4}, \frac{3\pi}{4}$ & $0, \pi$ & $0, \pi$ &  $\frac{\pi}{2}\pm\delta, \frac{3\pi}{2}\pm\delta$ \\
    5 & 64 & $\frac{\pi}{2}$ & $\frac{\pi}{4}, \frac{3\pi}{4}$ & $\pm\frac{\pi}{4}, \pm\frac{3\pi}{4}$ & $0, \pi$ & $0\pm\delta, \pi\pm\delta$  \\
    6 & 64 & $\frac{\pi}{4}, \frac{3\pi}{4}$ & $\frac{\pi}{2}$ & $\frac{\pi}{2}, \frac{3\pi}{2}$ & $\pm\frac{\pi}{4}, \pm\frac{3\pi}{4}$ &  $0\pm\delta, \pi\pm\delta$ \\
    7 & 64 & $\frac{\pi}{4}, \frac{3\pi}{4}$ & $\frac{\pi}{2}$ & $0, \pi$ & $\pm\frac{\pi}{4}, \pm\frac{3\pi}{4}$ & $0\pm\delta,\pi\pm\delta$ \\
    8 & 64 & $\frac{\pi}{2}$ & $\frac{\pi}{2}$ & $\pm\frac{\pi}{4}, \pm\frac{3\pi}{4}$ & $\pm\frac{\pi}{4}, \pm\frac{3\pi}{4}$ & $\frac{\pi}{2}\pm\delta, \frac{3\pi}{2}\pm\delta$ 
    \end{tabular}
    \end{ruledtabular}
    \captionof{table}{\label{tab:ED}%
    The parameters of the states with maximal magic on the ED boundary. The states are numbered from 0 to 9 in the order in which they appear in Fig.~\ref{fig:upper_pareto_right}. 
    }
\end{figure}

The right branch ED is simply a continuation of the middle branch GFE, along which the angle $\gamma$ is not constant, but depends on the concurrence $\Delta$. The values of the remaining four angles, $\theta_1$, $\theta_2$, $\phi_1$ and $\phi_2$, remain the same as in Table~\ref{tab:GFE}. The patterns of expectation values for states along the ED boundary and the corresponding parameter values are shown in Figure~\ref{fig:upper_pareto_right} and Table~\ref{tab:ED}, respectively. Figure~\ref{fig:upper_pareto_right} is very similar to Figure~\ref{fig:upper_pareto_middle}, the only difference is that the rightmost white square in each row of Figure~\ref{fig:upper_pareto_middle} has turned gray here, indicating an additional non-zero term. 
Analogously, Table~\ref{tab:ED} is similar to Table~\ref{tab:GFE}, the only difference being that the values for the $\gamma$ angles are shifted from their nominal values appearing in Table~\ref{tab:GFE} by a $\Delta$-dependent quantity $\delta(\Delta)$ that we derive below.

\begin{figure}[t]
\includegraphics[width=0.6\textwidth]{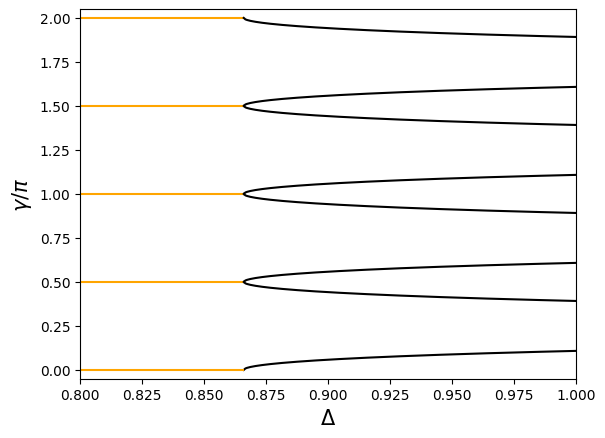}
\caption{\label{fig:gamma_right} The variation of the recurrence $\gamma$ (in units of $\pi$) along the boundary FED. The split occurs at $\Delta=\sqrt{\frac{3}{4}}$, see text.}
\end{figure}

To obtain $\delta$, we can follow the derivation of (\ref{eq:M2max_GFE_chi}), but revert one step and keep the general dependence on the angle $\gamma$:
\begin{subequations}
\begin{eqnarray}
\left| \langle \psi \vert I\otimes I \vert \psi \rangle \right|^4
&=& 1, \\[2mm]
  \left| \langle \psi \vert Y\otimes Y \vert \psi \rangle \right|^4
= \left| \langle \psi \vert Z\otimes Z \vert \psi \rangle \right|^4 
&=& \frac{1}{16}(1+\sin\chi\cos\gamma)^4 \\ [2mm]
  \left| \langle \psi \vert Y\otimes Z \vert \psi \rangle \right|^4 
= \left| \langle \psi \vert Z\otimes Y \vert \psi \rangle \right|^4 
&=& \frac{1}{16}(1-\sin\chi\cos\gamma)^4, \\[2mm]
  \left| \langle \psi \vert Y\otimes I \vert \psi \rangle \right|^4
= \left| \langle \psi \vert Z\otimes I \vert \psi \rangle \right|^4 &=&
\nonumber \\ [2mm]
= \left| \langle \psi \vert I\otimes Y \vert \psi \rangle \right|^4 
= \left| \langle \psi \vert I\otimes Z \vert \psi \rangle \right|^4 
&=& \frac{1}{4}\cos^4\chi, \\[2mm]
  \left| \langle \psi \vert X\otimes Y \vert \psi \rangle \right|^4
= \left| \langle \psi \vert Y\otimes X \vert \psi \rangle \right|^4 
&=& \nonumber\\ [2mm]
= \left| \langle \psi \vert X\otimes Z \vert \psi \rangle \right|^4 
= \left| \langle \psi \vert Z\otimes X \vert \psi \rangle \right|^4 
&=& \frac{1}{4}\sin^4\chi\sin^4\gamma, \\ [2mm]
  \left| \langle \psi \vert X\otimes X \vert \psi \rangle \right|^4 
&=& \sin^4\chi \cos^4\gamma, 
\end{eqnarray}
\label{eq:contributions_ED}
\end{subequations}
As a cross-check, we can verify that eqs.~(\ref{eq:contributions_ED}) reduce to eqs.~(\ref{eq:contributions_GFE}) for $\cos\gamma=0$. Next, we substitute (\ref{eq:contributions_ED}) in (\ref{eq:M2def}) to obtain
\begin{equation}
M_2(\chi,\gamma)= - \ln \left(\frac{\frac{5}{4}+\cos^4\chi
+ \sin^4\chi(\sin^4\gamma+\frac{5}{4}\cos^4\gamma) +\frac{3}{2}\sin^2\chi\cos^2\gamma}{4}\right).
\label{eq:M2_gamma}
\end{equation}
It is easy to show that this expression is maximized for
\begin{equation}
\cos^2\gamma = \frac{4\sin^2\chi-3}{9\sin^2\chi} \qquad \Leftrightarrow \qquad 
\sin^2\gamma = \frac{5\sin^2\chi+3}{9\sin^2\chi}.
\label{eq:gamma_solutions_at_max}
\end{equation}
Since the nominal value for $\gamma$ for this case was $\gamma=\frac{\pi}{2}$ (see the 0-th row in Table~\ref{tab:GFE}), we conclude that the change in $\gamma$ from its nominal value is given by
\begin{equation}
\delta(\Delta) = \frac{\pi}{2} - \arccos{\sqrt{\frac{4\sin^2\chi-3}{9\sin^2\chi}}} = \frac{\pi}{2} - \arccos{\sqrt{\frac{4-3/\Delta^2}{9}}}.
\label{eq:delta_gamma}
\end{equation}
Notice that this expression is well-defined only for $\Delta\ge \sqrt{\frac{3}{4}}$.
At $\Delta=\sqrt{\frac{3}{4}}$ (which is the precise location of point E), the boundary ED merges tangent-continuously into the boundary GFE (see eq.~(\ref{eq:GFE_ED_merger})).
The bifurcation at point E in the values of $\gamma$ due to the additional contribution~(\ref{eq:delta_gamma}) is illustrated in Figure~\ref{fig:gamma_right}. The orange lines represent the values of the recurrence angle $\gamma$ (in units of $\pi$) along the GFE boundary, while the black lines give the values of $\gamma$ (again, in units of $\pi$) along the ED boundary. 


Finally, substituting (\ref{eq:gamma_solutions_at_max}) in 
(\ref{eq:M2_gamma}), trading $\sin\chi$ for $\Delta$, and simplifying, we find
\begin{equation}
M_2^{(\text{max})}(\Delta)
\equiv f_\text{ED}(\Delta)
= 
\ln \left(\frac{18}{7\Delta^4 - 6\Delta^2 +9}\right),
\label{eq:M2max_ED_Delta} 
\end{equation}
which is the result in (\ref{eq:fED}).

This completes our discussion of the maximal magic boundary IHGFED. Before we conclude, we point out another curious fact: at $\Delta=1$, both $f_\text{ED}$ and $f_\text{IHG}$ give an identical answer:
\begin{equation}
M_2^{(\text{max})}(\Delta=1) = f_\text{ED}(\Delta=1) = f_\text{IHG}(\Delta=1) = \ln\left(\frac{9}{5}\right).
\end{equation}
In other words, after crossing the extension of the GFE boundary at $\Delta=0.977411$, the extension of the IHG boundary eventually meets the ED boundary at point D at $\Delta=1$ (see also Figures~\ref{fig:M2_versus_Delta} and \ref{fig:paretos}).

\section{Summary and Discussion}
\label{sec:conclusions}

\begin{figure}[t]
\includegraphics[width=0.9\textwidth]{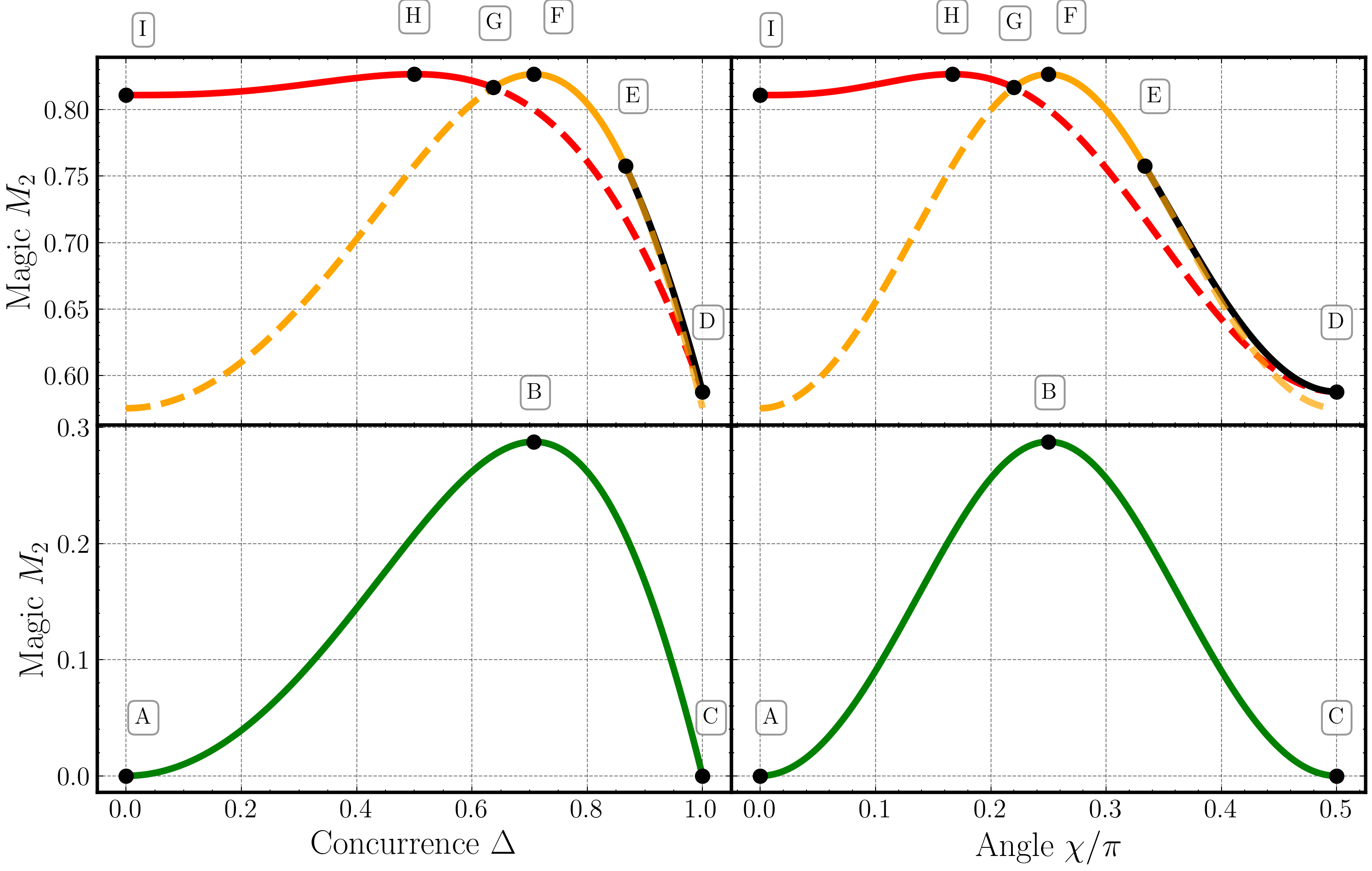}
\caption{\label{fig:paretos} The Pareto frontiers discussed in this paper: $f_\text{ABC}$ (lower panels, green line), 
$f_\text{IHG}$ (upper panels, red line),
$f_\text{GFE}$ (upper panels, orange line), and
$f_\text{ED}$ (upper panels, black line). In the left panels the magic is plotted versus the concurrence $\Delta$, while in the right panels it is plotted versus the concurrence angle $\chi$ in units of $\pi$. Points A through I denote the interesting features exhibited in Figure~\ref{fig:M2_versus_Delta}.
}
\end{figure}

In this paper we studied pure two-qubit states from the point of view of resource theory. The two relevant resources were the magic, quantified by the stabilizer R\'enyi entropy of order two, $M_2$, and the entanglement, characterized by the concurrence~$\Delta$. We showed that for pure two-qubit states with fixed concurrence $\Delta$, the magic measure $M_2$ always lies in the interval
\begin{equation}
M_2^{\text{(min)}}(\Delta) \le M_2(\Delta) \le M_2^{\text{(max)}}(\Delta)\,.
\end{equation}
We derived analytical expressions for the two boundaries $M_2^{\text{(min)}}(\Delta)$ and $M_2^{\text{(max)}}(\Delta)$, which are summarized in eqs.~(\ref{eq:M2min_delta}-\ref{eq:f_functions}). Figure~\ref{fig:paretos} illustrates those boundaries.
In particular, we find that the boundary $M_2^{\text{(max)}}(\Delta)$ is piecewise-defined, and consists of three segments: IHG, GFE and ED, discussed in Section~\ref{sec:IHG}, Section~\ref{sec:GFE}, and Section~\ref{sec:ED}, respectively.

We identified the set of quantum states on each boundary line: we found 144 states on the lower Pareto boundary ABC (listed in Table~\ref{tab:ABC}), 192 states on the upper Pareto boundary segment IHG (see Table~\ref{tab:IHG}), 288 states on the upper Pareto boundary segment GFE (see Table~\ref{tab:GFE}) and 576 states on the upper Pareto boundary segment ED (see Table~\ref{tab:ED}). In particular, we confirm the result from Ref.~\cite{Liu:2025frx} that there are a total of 480 states with maximal magic of $M_2=\ln\left(\frac{16}{7}\right)$. Of those states, 288 have concurrence $\Delta=1/\sqrt{2}$ and 192 have concurrence $\Delta=1/2$. This can be easily verified by taking any one of those maximal magic states and acting on it with the Clifford operators, which do not change the value of the magic. The result is a set of 192 distinct states at $\Delta_\text{H}=1/2$ and 288 distinct states at $\Delta_\text{F}=1/\sqrt{2}$. Repeating the same exercise for a state on the ED boundary, we find that the 576 states are divided into two classes of 288 states each, where each class belongs to a different Clifford orbit.

The non-trivial shapes of the Pareto frontiers IHGFED and ABC imply that magic and entanglement are complementary resources. Some noteworthy features are:
\begin{itemize}
\item The boundary IHG is relatively flat. For example, $M_2$ changes from $\ln\frac{9}{4}$ at point I to $\ln\frac{16}{7}$ at point H, which is a difference of less than 2\%. 
\item Maximal magic is not achieved at maximum entanglement, but at two intermediate values of the concurrence: $\Delta=1/2$ (point H) and $\Delta=1/\sqrt{2}$ (point F). 
\item The boundary FED represents a true entanglement-magic trade-off. Along that boundary, increasing the entanglement lowers the maximum possible magic and vice versa. 
\item The lower boundary ABC implies that partially entangled states are always magical. 
\item A close inspection of Figure~\ref{fig:paretos} reveals that the boundaries ABC and GFE have very similar shapes. This can be verified from our analytical expressions: the formula for the GFE boundary (\ref{eq:M2max_GFE_Delta}) can be equivalently rewritten as
$$
f_\text{GFE} (\Delta) = \ln 2 - \ln \left(\Delta^4-\Delta^2+1 + \frac{1}{8}\right).
$$
Comparing to eq.~(\ref{eq:M2min_delta_2}), we notice that the second term would have been exactly $f_\text{ABC}(\Delta)$ if not for the extra term $\frac{1}{8}$ in the argument of the logarithm. This also means that the ABC and GFE boundaries are shifted vertically from each other by approximately $\ln 2$ and that their peaks appear in the same location.
\item Typical states (the yellow-colored region in Figure~\ref{fig:M2_versus_Delta}) are both moderately entangled and moderately magical, and sit away from the Pareto boundaries.
\end{itemize}

\begin{acknowledgments}
We thank G.~Fleming, K.~Kong, H.~Lamm, T.~Menzo and P.~Vander Griend for useful discussions.
The work of AR, MK and KTM is supported in part by the Shelby Endowment for Distinguished Faculty at the University of Alabama. The work of AR and KM is supported in part by the Shelby Endowment for Distinguished Faculty at the University of Alabama and by Fermilab via Subcontract 731293, in support of DOE Award No.\ DE-SCL0000090 ``HEP AmSC IDA Pilot: Knowledge Extraction'' and DOE Award No.\ DE-SCL0000152 ``USQCD AmSC Infrastructure Provision''. The work of KM and SG is supported in part by the U.S. Department of Energy (DOE) under Award No. DE-SC0026347. The work of SG is supported in part by DOE under Award No. DE-SC0012447. 
\end{acknowledgments}

\appendix

\section{Catalog of the States on the IHG Boundary}
\label{app:catalog}

\begin{table}[t]
\caption{\label{tab:IHG14} An alternative version of Table~\ref{tab:IHG} where the states are listed in the order in which they appear in Figure~\ref{fig:upper_pareto_left}. The number in the first column represents the row in Figure~\ref{fig:upper_pareto_left}, and the different cases are indicated with lowercase Latin letters. Note that all angles are treated modulo $2\pi$, e.g., $\phi_1=-\frac{\pi}{4}$ and $\phi_1=\frac{7\pi}{4}$ are equivalent. Due to space shortage, the states from the last two rows in Figure~\ref{fig:upper_pareto_left} are listed separately in Table~\ref{tab:IHG56}.}
\begin{ruledtabular}
\scalebox{0.8}{
    \begin{tabular}{rcccccc}
    \small
    \textrm{No}&
    \textrm{Multiplicity}&
    $\theta_1$&
    $\theta_2$&
    $\phi_1$ &
    $\phi_2$ &
    $\gamma$\\
    \colrule
    0a & 16 & $\alpha,\ \pi-\alpha$ & $\theta_1$ & $\pm\frac{\pi}{4},\ \pm\frac{3\pi}{4}$ & $\phi_1\pm\frac{\pi}{2}$ & $\pi$ \\
    0b & 16 & $\alpha,\ \pi-\alpha$ & $\pi-\theta_1$ & $\pm\frac{\pi}{4},\ \pm\frac{3\pi}{4}$ & $\phi_1,\ \phi_1+\pi$ & $0$ \\
    1a & 8 & $\alpha$ & $\alpha$ & $\pm\frac{\pi}{4},\ \pm\frac{3\pi}{4}$ & $\phi_1,\ \phi_1+\pi$ &
    $\begin{cases}
    \frac{5\pi}{3}, & \phi_1=\frac{\pi}{4},\ \frac{5\pi}{4} \\
    \frac{\pi}{3}, & \phi_1=\frac{3\pi}{4},\ \frac{7\pi}{4}
    \end{cases}$ \\
    1b & 8 & $\pi-\alpha$ & $\pi-\alpha$ & $\pm\frac{\pi}{4},\ \pm\frac{3\pi}{4}$ & $\phi_1,\ \phi_1+\pi$ &
    $\begin{cases}
    \frac{\pi}{3}, & \phi_1=\frac{\pi}{4},\ \frac{5\pi}{4} \\
    \frac{5\pi}{3}, & \phi_1=\frac{3\pi}{4},\ \frac{7\pi}{4}
    \end{cases}$ \\
    1c &  8 & $\alpha$ & $\pi-\alpha$ & $\pm\frac{\pi}{4},\ \pm\frac{3\pi}{4}$ & $\phi_1\pm\frac{\pi}{2}$ &
    $\begin{cases}
    \frac{2\pi}{3}, & \phi_1=\frac{\pi}{4},\ \frac{5\pi}{4} \\
    \frac{4\pi}{3}, & \phi_1=\frac{3\pi}{4},\ \frac{7\pi}{4}
    \end{cases}$ \\
    1d &  8 & $\pi-\alpha$ & $\alpha$ & $\pm\frac{\pi}{4},\ \pm\frac{3\pi}{4}$ & $\phi_1\pm\frac{\pi}{2}$ &
    $\begin{cases}
    \frac{4\pi}{3}, & \phi_1=\frac{\pi}{4},\ \frac{5\pi}{4} \\
    \frac{2\pi}{3}, & \phi_1=\frac{3\pi}{4},\ \frac{7\pi}{4}
    \end{cases}$ \\
    2a &  16 & $\alpha,\ \pi-\alpha$ & $\theta_1$ & $\pm\frac{\pi}{4},\ \pm\frac{3\pi}{4}$ & $\phi_1,\ \phi_1+\pi$ & $\pi$ \\
    2b &  16 & $\alpha,\ \pi-\alpha$ & $\pi-\theta_1$ & $\pm\frac{\pi}{4},\ \pm\frac{3\pi}{4}$ & $\phi_1\pm\frac{\pi}{2}$ & $0$ \\
    3a &  8 & $\alpha$ & $\alpha$ & $\pm \frac{\pi}{4},\ \pm\frac{3\pi}{4}$ & $\phi_1\pm\frac{\pi}{2}$ & $
    \begin{cases}
    \frac{5\pi}{3}, & \phi_1=\frac{\pi}{4},\ \frac{5\pi}{4} \\
    \frac{\pi}{3}, & \phi_1=\frac{3\pi}{4},\ \frac{7\pi}{4}
    \end{cases} $\\
    3b &  8 & $\pi-\alpha$ & $\pi-\alpha$ & $\pm\frac{\pi}{4},\ \pm\frac{3\pi}{4}$ & $\phi_1\pm\frac{\pi}{2}$ & $
    \begin{cases}
    \frac{\pi}{3}, & \phi_1=\frac{\pi}{4},\ \frac{5\pi}{4} \\
    \frac{5\pi}{3}, & \phi_1=\frac{3\pi}{4},\ \frac{7\pi}{4}
    \end{cases} $\\
    3c & 8 & $\alpha$ & $\pi-\alpha$ & $\pm\frac{\pi}{4},\ \pm\frac{3\pi}{4}$ & $\phi_1,\ \phi_1+\pi$ &$
    \begin{cases}
    \frac{2\pi}{3}, & \phi_1=\frac{\pi}{4},\ \frac{5\pi}{4} \\
    \frac{4\pi}{3}, & \phi_1=\frac{3\pi}{4},\ \frac{7\pi}{4}
    \end{cases} $\\
    3d & 8 & $\pi-\alpha$ & $\alpha$ & $\pm\frac{\pi}{4},\ \pm\frac{3\pi}{4}$ & $\phi_1,\ \phi_1+\pi$ &$
    \begin{cases}
    \frac{4\pi}{3}, & \phi_1=\frac{\pi}{4},\ \frac{5\pi}{4} \\
    \frac{2\pi}{3}, & \phi_1=\frac{3\pi}{4},\ \frac{7\pi}{4}
    \end{cases} $
    \end{tabular}
}
\end{ruledtabular}
\end{table}

In Section~\ref{sec:IHG} we showed that the states on the IHG boundary give rise to six distinct patterns in the magic sum which were depicted in Figure~\ref{fig:upper_pareto_left}. Table~\ref{tab:IHG} listed all 192 of those states in a rather compact notation that takes only four rows, but hides the correspondence of the state to the patterns in Figure~\ref{fig:upper_pareto_left}. In this appendix we rewrite Table~\ref{tab:IHG} so that the correspondence to the pattern in Figure~\ref{fig:upper_pareto_left} is explicit. The result is rather lengthy, which is why it is spread over two tables, Table~\ref{tab:IHG14} and \ref{tab:IHG56}. In each of those, the number in the first column represents the row in Figure~\ref{fig:upper_pareto_left}, within which different cases are indicated with lowercase Latin letters. 

\begin{table}
\caption{\label{tab:IHG56} Continuation of Table~\ref{tab:IHG14} listing the states from the last two rows in Figure~\ref{fig:upper_pareto_middle}.}
\begin{ruledtabular}
\scalebox{0.8}{
    \begin{tabular}{rccccccc}
    \small
    \textrm{No}&
    \textrm{Multiplicity}&
    $\theta_1$&
    $\theta_2$&
    $\phi_1$ &
    $\phi_2$ &
    $\gamma$\\
    \colrule
    4a &  8 & $\alpha$ & $\alpha$ & $\pm\frac{\pi}{4},\ \pm\frac{3\pi}{4}$ & $\phi_1\pm\frac{\pi}{2}$ &
    $\begin{cases}
    \frac{\pi}{3}, & \phi_1=\frac{\pi}{4},\ \frac{5\pi}{4} \\
    \frac{5\pi}{3}, & \phi_1=\frac{3\pi}{4},\ \frac{7\pi}{4}
    \end{cases}$ \\
    4b &  8 & $\pi-\alpha$ & $\pi-\alpha$ & $\pm\frac{\pi}{4},\ \pm\frac{3\pi}{4}$ & $\phi_1\pm\frac{\pi}{2}$ &
    $\begin{cases}
    \frac{5\pi}{3}, & \phi_1=\frac{\pi}{4},\ \frac{5\pi}{4} \\
    \frac{\pi}{3}, & \phi_1=\frac{3\pi}{4},\ \frac{7\pi}{4}
    \end{cases}$ \\
    4c &  8 & $\alpha$ & $\pi-\alpha$ & $\pm\frac{\pi}{4},\ \pm\frac{3\pi}{4}$ & $\phi_1,\ \phi_1+\pi$ &
    $\begin{cases}
    \frac{4\pi}{3}, & \phi_1=\frac{\pi}{4},\ \frac{5\pi}{4} \\
    \frac{2\pi}{3}, & \phi_1=\frac{3\pi}{4},\ \frac{7\pi}{4}
    \end{cases}$ \\
    4d & 8 & $\pi-\alpha$ & $\alpha$ & $\pm\frac{\pi}{4},\ \pm\frac{3\pi}{4}$ & $\phi_1,\ \phi_1+\pi$ &
    $\begin{cases}
    \frac{2\pi}{3}, & \phi_1=\frac{\pi}{4},\ \frac{5\pi}{4} \\
    \frac{4\pi}{3}, & \phi_1=\frac{3\pi}{4},\ \frac{7\pi}{4}
    \end{cases}$ \\    
    5a & 8 & $\alpha$ & $\alpha$ & $\pm\frac{\pi}{4},\ \pm\frac{3\pi}{4}$ & $\phi_1,\ \phi_1+\pi$ &
    $\begin{cases}
    \frac{\pi}{3}, & \phi_1=\frac{\pi}{4},\ \frac{5\pi}{4} \\
    \frac{5\pi}{3}, & \phi_1=\frac{3\pi}{4},\ \frac{7\pi}{4}
    \end{cases}$ \\
    5b & 8 & $\pi-\alpha$ & $\pi-\alpha$ & $\pm\frac{\pi}{4},\ \pm\frac{3\pi}{4}$ & $\phi_1,\ \phi_1+\pi$ &
    $\begin{cases}
    \frac{5\pi}{3}, & \phi_1=\frac{\pi}{4},\ \frac{5\pi}{4} \\
    \frac{\pi}{3}, & \phi_1=\frac{3\pi}{4},\ \frac{7\pi}{4}
    \end{cases}$ \\
    5c & 8 & $\alpha$ & $\pi-\alpha$ & $\pm\frac{\pi}{4},\ \pm\frac{3\pi}{4}$ & $\phi_1\pm\frac{\pi}{2}$ &
    $\begin{cases}
    \frac{4\pi}{3}, & \phi_1=\frac{\pi}{4},\ \frac{5\pi}{4} \\
    \frac{2\pi}{3}, & \phi_1=\frac{3\pi}{4},\ \frac{7\pi}{4}
    \end{cases}$ \\
    5d & 8 & $\pi-\alpha$ & $\alpha$ & $\pm\frac{\pi}{4},\ \pm\frac{3\pi}{4}$ & $\phi_1\pm\frac{\pi}{2}$ &
    $\begin{cases}
    \frac{2\pi}{3}, & \phi_1=\frac{\pi}{4},\ \frac{5\pi}{4} \\
    \frac{4\pi}{3}, & \phi_1=\frac{3\pi}{4},\ \frac{7\pi}{4}
    \end{cases}$ \\
    \end{tabular}
}
\end{ruledtabular}
\end{table}

Tables~\ref{tab:IHG14} and \ref{tab:IHG56} reveal that each row of Figure~\ref{fig:upper_pareto_left} contains 32 states, so that the states in all six rows in Figure~\ref{fig:upper_pareto_left} indeed add up to the 192 states depicted in the original Table~\ref{tab:IHG} from Section~\ref{sec:IHG}.

\bibliography{refs_arxiv}

\end{document}